\documentclass[fleqn,usenatbib]{mnras}
\usepackage[T1]{fontenc}
\usepackage{ae,aecompl}
\usepackage{graphicx}
\usepackage{amsmath}
\usepackage{amssymb}
\usepackage{txfonts}
\usepackage{cleveref}
\usepackage{bm}
\usepackage{hyperref}
\usepackage{ upgreek }
\usepackage{tensor}
\usepackage{mathtools}

\usepackage{cleveref}
\usepackage{upgreek}
\usepackage{xcolor}

\graphicspath{./}

\newcommand{\ltsima}{$\; \buildrel < \over \sim \;$}
\newcommand{\lsim}{\lower.5ex\hbox{\ltsima}}
\newcommand{\gtsima}{$\; \buildrel > \over \sim \;$}
\newcommand{\gsim}{\lower.5ex\hbox{\gtsima}}

\newcommand{\chip}{{\chi^\prime}}

\newcommand{\revision}[1]{\textcolor{black}{#1}}

\title[functional photo-$z$]
{Propagating photo-$z$ uncertainties: a functional derivative approach}
\author[Reischke]
{
Robert Reischke\thanks{E-mail:  \href{mailto:reischke@astro.ruhr-uni-bochum.de}{reischke@astro.ruhr-uni-bochum.de}}$^{1}$
\\
$^1$ Ruhr University Bochum, Faculty of Physics and Astronomy, Astronomical Institute (AIRUB),\\ \hspace{0.15cm} German Centre for Cosmological Lensing, 44780 Bochum, Germany
}

\begin{document}
 \pagerange{\pageref{firstpage}--\pageref{lastpage}}
\pubyear{2022}
\maketitle
\label{firstpage}

\begin{abstract}
Photometric redshifts are a key ingredient in the analysis and interpretation of large-scale structure (LSS) surveys. The accuracy and precision of these redshift estimates are directly linked to the constraining power of photometric surveys. It is hence necessary to define precision and accuracy requirements for the redshift calibration \revision{to not} infer biased results in the final analysis. 
For weak gravitational lensing of the LSS, the photometry culminates in the estimation of the source redshift distribution (SRD) in each of the tomographic bins used in the analysis. The focus has been on shifts of the mean of the SRDs and how well the calibration must be able to recover those. Since the estimated SRDs are usually given as a normalized histogram with corresponding errors, it would be advantageous to propagate these uncertainties accordingly to see whether the requirements of the given survey are indeed fulfilled. Here we propose the use of functional derivatives to calculate the sensitivity of the final observables, e.g. the lensing angular power spectrum, with respect to the SRD at a specific redshift. This allows the propagation of arbitrarily shaped small perturbations to the SRD, without having to run the whole analysis pipeline for each realization again. We apply our method to an EUCLID survey and demonstrate it with SRDs of the KV450 data set, recovering previous results. \revision{Lastly, we note that the moments of the SRD of order larger than two will probably not be relevant when propagating redshift uncertainties in cosmic shear analysis.}
\end{abstract}


\section{Introduction}
\label{sec:intro}
{\revision{
Cosmic shear, the weak gravitational lensing effect imprinted on distant galaxies by the large-scale structure (LSS), is one of the primal science goals for EUCLID and Rubin-LSST. The blueprint for these missions has been set by current stage-3 surveys, including the Kilo-Degree Survey \citep[KiDS]{hildebrandt_kids-450_2017,asgari_kids-1000_2021}\footnote{\href{https://kids.strw.leidenuniv.nl/}{https://kids.strw.leidenuniv.nl/}}, the Dark Energy Survey \citep[DES]{abbott_dark_2018,2022PhRvD.105b3514A}\footnote{\href{https://www.darkenergysurvey.org/}{https://www.darkenergysurvey.org/}} and the Subaru Hyper Suprime-Cam \citep[HSC]{takada_subaru_2010, hamana_cosmological_2020}\footnote{\href{https://hsc.mtk.nao.ac.jp/ssp/}{https://hsc.mtk.nao.ac.jp/ssp/}}, yielding tight constraints on the matter distribution in the late Universe.}}

The cosmic shear signal is estimated by measuring the coherent distortion of background galaxies. Since the intrinsic ellipticity of galaxies is much larger than the lensing effect, millions (or even billions) of galaxies are required to measure a significant signal. This makes a complete spectroscopic survey unfeasible. 
{\revision{There are two main techniques to obtain an estimate of the true redshift for the background galaxy sample: The first one is called photometric redshifts (see e.g. \citealt{lima_estimating_2008, bonnett_redshift_2016, hildebrandt_kids-1000_2020}), that is using broadband photometry which is then calibrated using a significantly smaller spectroscopic reference sample. Recently self-organising maps have become a standard technique for photometric redshifts \citep{masters_mapping_2015,wright_photometric_2020,hildebrandt_kids-1000_2020,myles_dark_2021}. The second approach is clustering redshifts (see e.g. \citealt{newman_calibrating_2008,matthews_reconstructing_2010} or \citealt{busch_testing_2020,2022MNRAS.510.1223G} for more recent works), where the redshift distribution of the galaxy sample is estimated by an angular cross-correlation measurement. There also exist hybrid methods, combining photometry and clustering measurements into a Bayesian hierarchical model \citep{sanchez_redshift_2019,alarcon_redshift_2020,rau_estimating_2020, 2022MNRAS.509.4886R}.}}

{\revision{All of the above methods yield an as unbiased as possible estimate of the distribution of background galaxies in redshift (source-redshift-distribution, SRD). The statistical precision of the experiments sets limits on the required accuracy of the SRD estimation technique. Likewise, the question arises of how the residual uncertainties in the SRD itself affect the inference process. Most works use simple shift parameters in the mean of the SRD which are then marginalised over in a given (informed) prior range. There are, however, other approaches using different parametrisation for these uncertainties which itself can be informative priors or even be self-calibrated using different two-point function measurements. This includes non-parametric approaches \citep{rau_estimating_2020}, using higher moments beyond the mean \citep{mcleod_joint_2017}, explicit parametrisation of outliers in photometric redshifts \citep{schaan_photo-z_2020} or Gaussian mixture models \citep{stolzner_self-calibration_2021}
}}

{\revision{Recently full shape techniques, i.e. methods beyond the mean shifts of the SRD, have been applied to real data. In \citet{2022PhRvD.105b3514A} used \texttt{Hyperrank} \citep{2022MNRAS.511.2170C} to propagate the uncertainties of the SRD, finding that simple shift parameters are sufficient for DES. For the next generation of surveys, stage-4, this might no longer be true, however. Including the full-shape SRD might also require more efficient techniques for the marginalisation over the nuisance parameters, something which is for example analytically done in \citet{stolzner_self-calibration_2021} thanks to the Gaussian mixture model used. Another possibility was presented in \citet{2023MNRAS.518..709Z} where the SRD was sampled from the full shape and the resulting Markov chains of the cosmological parameters were combined using Baysian model evidence.}}

{\revision{In summary, it is not entirely clear to date how accurate the error propagation of the residual SRD uncertainties have to be in order to obtain unbiased cosmological results, i.e. uncertainties on the cosmological parameters together with their maximum posterior values. While unbiased results are obtained with simple shifts at the moment, EUCLID and Rubin-LSST will change this and the full shape of the SRD posterior distribution becomes important for cosmological inference. It is hence vital to investigate the most general sensitivity of cosmic shear observables to the underlying SRD, without assuming a specific functional form of the SRD itself. Therefore, in this paper, we calculate the functional derivative of the cosmic shear angular power spectrum with respect to the SRD. That is, we investigate arbitrary (but small) perturbations to the SRD at a certain redshift and how they propagate into the angular power spectrum. This functional derivative can then be used to calculate the total error in the measurement process by simple error propagation. We take the constraint of the normalisation of the SRD into account when calculating the functional derivative. Therefore we can propagate arbitrary perturbations to the SRDs (subject to some underlying covariance) and propagate them into the $C_\ell$ of cosmic shear. This allows us to estimate the difference in $\chi^2$ induced by the uncertainty in the SRD, without having to run thousands of realizations of the analysis pipeline used. 
By using a Fisher matrix for the cosmological parameters, this $\Delta\chi^2$ can then be mapped to potential biases in cosmological parameters. Here we studied a rather idealised scenario by working in Fourier space, assuming a Gaussian likelihood and ignoring intrinsic alignments. The method, however, easily generalises and including these effects is straightforward.
The approach therefore tries to fill the gap between cheap marginalisation over shift parameters and very expensive marginalisation over different Monte Carlo Markov Chains (MCMCs). It furthermore allows us to investigate where uncertainty in the SRD is wreaking the most havoc on cosmological inference.}}

We structure the paper as follows: In \Cref{sec:methodology} we briefly review cosmic shear basics and introduce the methodology used by calculating the functional derivative of the weak lensing angular power spectrum. The results are presented in \Cref{sec_results}, where we apply the procedure to a survey with EUCLID's specifications and to KiDS-VIKING-450 (KV450). We conclude in \Cref{sec:conclusions}. In the appendices, we also investigate the possibility of an Edgeworth expansion of the SRD (\Cref{ssec:edgeworth expansion}), discuss photometric galaxy clustering (\Cref{sec:photometric_clustering}), the distribution of the mean and standard deviation of the SRD in \Cref{sec:m_and_v_kv450},
  the general relationship to observables (\Cref{sec:observables}), the functional derivative of the non-Limber projection in \Cref{sec:nonlimber} and intrinsic alignments (\Cref{sec:intrinsics}).

\section{Methodology}
\label{sec:methodology}
In this section, we present the basic methodology of our analysis. In particular, we describe the basics of cosmic shear and derive the functional derivative of the lensing angular power spectrum with respect to the SRDs.

\subsection{Cosmic shear basics}
The equation for the cosmic shear power spectrum in tomographic bins $i$ and $j$ in the Limber projection is \citep{limber_analysis_1954,loverde_extended_2008}
\begin{equation}
\label{eq:limber}
    C^{\kappa_i\kappa_j}_\ell = \int_0^{\chi_\mathrm{H}}\frac{\mathrm{d}\chi}{\chi^2} W^{(i)}_\kappa(\chi)W^{(j)}_\kappa(\chi) P_\delta\left(\frac{\ell + 0.5}{\chi},\chi\right)\;,
\end{equation}
where $P_\delta$ is the matter power spectrum, for which we use the emulated spectrum from \citet{mead_accurate_2015}. $W^{(i)}_\kappa (\chi)$ is the lensing weight of the $i$-th tomographic bin as given by:
\begin{equation}
\label{eq:weight}
    W^{(i)}_\kappa(\chi) =\frac{3\Omega_\mathrm{m0}}{2\chi_\mathrm{H}^2}\frac{\chi}{a(\chi)}\int_\chi^{\chi_\mathrm{H}}\mathrm{d}\chip n^{(i)}_\mathrm{s}(\chip)\frac{\chip-\chi}{\chip}\;.
\end{equation}
Here $\chi$ is the co-moving distance, $a$ the scale factor, $\Omega_{\mathrm{m}0}$ the matter density parameter today, $\chi_\mathrm{H}$ the Hubble radius and $n^{(i)}_\mathrm{s}$ is the SRD in the $i$-th tomographic bin which builds on photo-$z$ measurements and its calibration. It is normalized in each tomographic bin such that
\begin{equation}
\label{eq:srd_norm}
    \int \mathrm{d}z\; n^{(i)}_\mathrm{s}(z) = 1 = \int\mathrm{d}\chi \;n^{(i)}_\mathrm{s}(z(\chi))\frac{\mathrm{d}z}{\mathrm{d}\chi}\equiv \int \mathrm{d}\chi \; n^{(i)}_\mathrm{s}(\chi)\;. 
\end{equation}
Since photo-$z$ is just an estimate of the true redshift, the estimated source-redshift distribution, $n^{(i)}_\mathrm{s}$, is not exactly known. 
Here we investigate two approaches:
\begin{enumerate}
    \item [i)] Use functional derivatives to evaluate the change of the lensing power spectrum when perturbing the $n^{(i)}_\mathrm{s}$ at different redshifts. Given specific survey settings and precision goals, limits on the allowed change of the $n^{(i)}_\mathrm{s}$ can be determined, which in turn can be mapped to changes in the cumulants or moments of the underlying distribution (see \Cref{ssec:functional_derivative}).
    \item [ii)]  We expand the underlying source-redshift distribution in an asymptotic Edgeworth series and investigate the requirements on the cumulants directly in a Fisher analysis. The second approach is not feasible for realistic SRDs (see \Cref{ssec:edgeworth expansion}).
\end{enumerate}

\subsection{Functional derivative of the lensing power spectrum}
\label{ssec:functional_derivative}
Here we wish to investigate the sensitivity of the weak lensing power spectrum to the full shape of the source-redshift distribution using functional derivatives. In particular we start by perturbing $n^{(i)}_\mathrm{s}(\chi(z))$ at a certain redshift $z_0$, such that $\chi_0 = \chi(z_0) $. The corresponding perturbed lensing weight is thus
\begin{equation}
\label{eq:delta_weight}
    \Delta W^{(i)}_\kappa(\chi, \chi_0) = \frac{\delta W^{(i)}_\kappa(\chi) }{\delta n^{(i)}_\mathrm{s}(\chi_0)}\Delta n^{(i)}_\mathrm{s}(\chi_0)\;.
\end{equation}
This expression evaluates, how the lensing weight changes if the source-redshift distribution is perturbed by an amount $\Delta n^{(i)}_\mathrm{s}$ at the co-moving distance $\chi_0$ corresponding to the redshift $z_0$.

Ultimately, we are interested in the change of the lensing power spectrum, \Cref{eq:limber}. First, by applying the Leibniz rule
\begin{equation}
\label{eq:derivative_cell}
\begin{split}
     \frac{\delta C^{(ij)}_\ell}{\delta n^{(a)}(\chi_0)} = & \;  \int\mathrm{d}x \frac{\delta C^{(ij)}_\ell}{\delta W^{(a)}(x)} \frac{\delta W^{(a)}(x)}{\delta n^{(a)}(\chi_0)} \\= & \;  \int\mathrm{d}x \frac{\delta W^{(a)}(x)}{\delta n^{(a)}(\chi_0)}\frac{P_\delta\left(\frac{\ell + 0.5}{x},x\right)}{x^2}\left(W^{(j)}(x)\delta^\mathrm{K}_{ia} + W^{(i)}(x)\delta^\mathrm{K}_{ja}\right)  \;,
\end{split}
\end{equation}
The missing ingredient is the functional derivative of the lensing kernel, for which we find
\begin{equation}
    \frac{\delta W^{(i)}(x) }{\delta n^{(j)}_\mathrm{s}(\chi_0)} =  \frac{3\Omega_\mathrm{m0}}{2\chi_\mathrm{H}^2}\frac{x}{a(x)} \frac{\chi_0-x}{\chi_0}\delta^\mathrm{K}_{ij}\Theta(\chi_0-x)\;.
\end{equation}
$\Theta(x)$ is the Heaviside function to ensure that the functional derivative vanishes if the SRD is perturbed outside the integration bounds.
Using \Cref{eq:delta_weight} and \Cref{eq:derivative_cell} we can write the change in angular power spectrum $\Delta C^{(ij)}_\ell (\chip)$ due to a change in the source-redshift distribution at co-moving distance $\chi_0$ as
\begin{equation}
\begin{split}
        \Delta C^{(ij)}_{\ell,a} (\chi_0) 
        \equiv & \ \frac{\delta C^{(ij)}_\ell}{\delta n^{(a)}(\chi_0)}  \Delta n^{(a)}(\chi_0) \\
        = &  \ \frac{3 \Omega_\mathrm{m0}}{2\chi_\mathrm{H}^2}\Delta n (\chi_0) \int  \frac{\mathrm{d}x}{a(x)x}\frac{\chi_0-x}{\chi_0} P_\delta\left(\frac{\ell + 0.5}{x},x\right)\\ &\times
       \left(W^{(j)}(x)\delta^\mathrm{D}_{ia} + W^{(i)}(x)\delta^\mathrm{D}_{ja}\right)\;.
        \end{split}
\end{equation}
Integrating the perturbed lensing spectrum then gives the total perturbation:
\begin{equation}
\label{eq:integrated_error}
     \Delta C^{(ij)}_{\ell,a} \equiv\int\mathrm{d}\chi_0 \Delta C^{(ij)}_{\ell,a} (\chi_0)\;.
\end{equation}
So far we have treated the function $n^{(i)}(z)$ as being completely free. However, the functional derivative needs to respect the constraint given in \Cref{eq:srd_norm}, thus limiting the possible variations of $n^{(i)}(z)$. The normalization condition itself is again a function and we write 
\begin{equation}
    N[n^{(i)}_\mathrm{s}]  \coloneqq 1 -\int \mathrm{d}z\; n^{(i)}_\mathrm{s}(z) = 0\;,
\end{equation}
this constraint can be implemented by first defining
\begin{equation}
    n^{(i)}_\mathrm{s}(z) \coloneqq\frac{f(z)}{\int \mathrm{d}x^\prime f(x^\prime)}
\end{equation}
which will be normalized by construction. $n^{(i)}_\mathrm{s}(z)$ is a functional of $f$ and we can now evaluate the functional derivative of $C[n[f]]$ as an unconstrained derivative but evaluated at $f=n$. To avoid clutter we ignore the sub- and superscripts in this part
\begin{equation}
    \left(\frac{\delta C[n[f]]}{\delta f(x)}\right)\Bigg|_{f=n} = \int \mathrm{d}x^\prime \frac{\delta C[n]}{\delta n(x^\prime)}\frac{\delta n(x^\prime)}{\delta f(x)}\Bigg|_{f=n}\;.
\end{equation}
With 
\begin{equation}
    \frac{\delta n(x^\prime)}{\delta f(x)} =  \frac{\delta_\mathrm{D}(x^\prime - x)}{\int\mathrm{d}y\;f(y)} - \frac{f(x^\prime)}{\left(\int\mathrm{d}y\;f(y)\right)^2}\;,
\end{equation}
one finds
\begin{equation}
\begin{split}
    \frac{\delta C[n]}{\delta_1 n(x)} \equiv & \ \left(\frac{\delta C[n[f]]}{\delta f(x)}\right)\Bigg|_{f=n} = \frac{\delta C[n]}{\delta n(x)} - \int \mathrm{d}y \; \frac{\delta C[n]}{\delta n(y)}n(y)\;,
\end{split}
\end{equation}
where we denote that we want to keep the normalization fixed by the variation $\delta_1$. This is a very intuitive expression: the first term evaluates the standard functional derivative, while the second term corrects this variation to respect the normalization. 

\begin{figure}
    \centering
    \includegraphics[width = .45\textwidth]{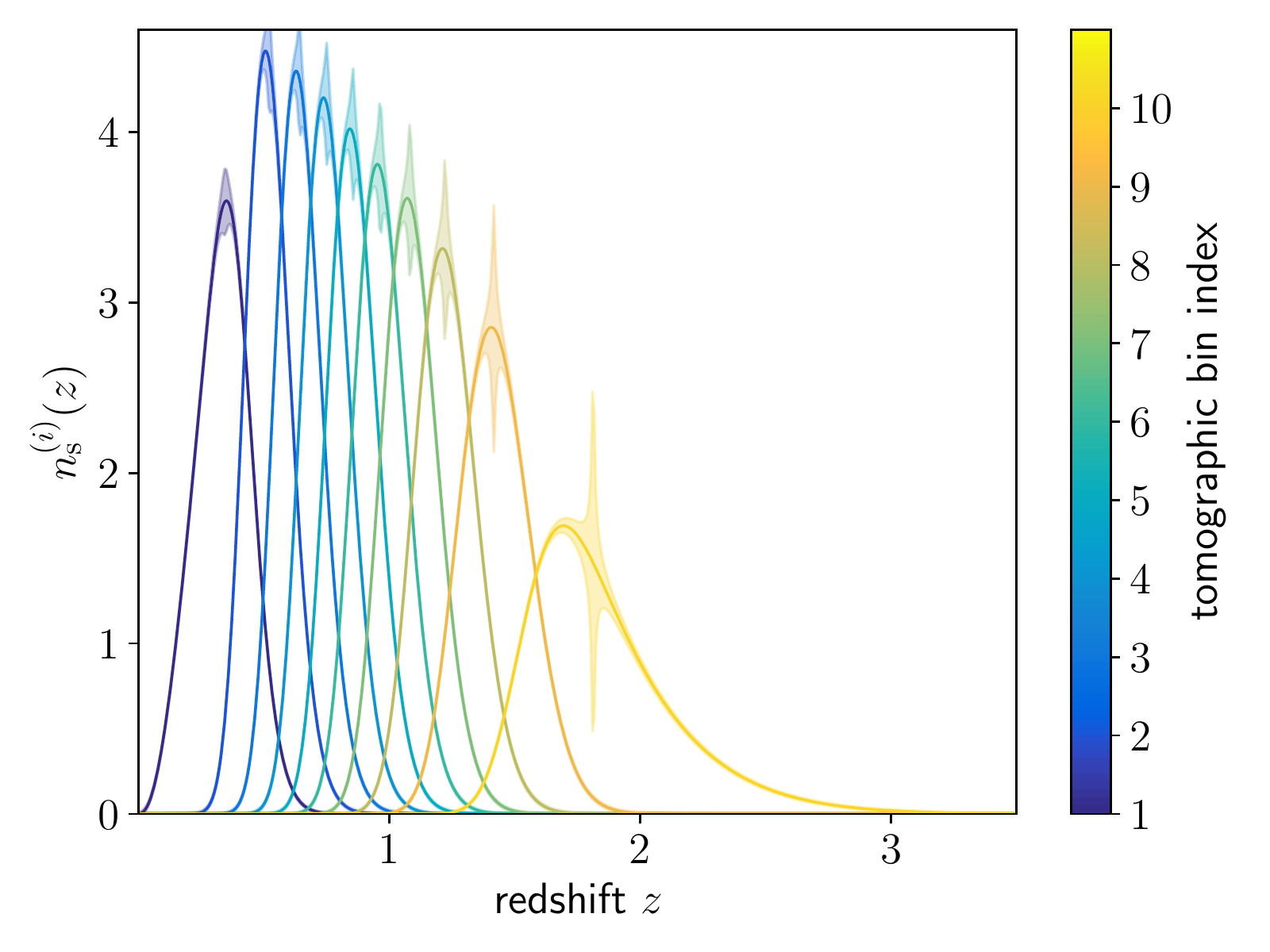}
    \caption{Allowed perturbation for EUCLID to the SRD of the ten tomographic source bins. Solid lines show the fiducial SRD, while the bands show the allowed perturbation to it, \revision{corresponding to a $\Delta\chi^2 = 1$.}}
    \label{fig:allowed_ecl}
\end{figure}

\subsection{Fisher forecast}
The next step is to set some requirements for the lensing power spectra. Here we will look at the difference in the $\chi^2$, assuming a Gaussian likelihood and thus setting a lower limit on the required accuracy of $n^{(i)}_\mathrm{s}(z)$. For modes $\boldsymbol{a}_{\ell m}$ with zero mean and covariance $\boldsymbol{C}_\ell$, the $\Delta\chi^2$ between multipoles $\ell_\mathrm{min}$ and $\ell_\mathrm{max}$ can be written as
\begin{equation}
\label{eq:delta_chi2}
    \Delta\chi^2 (\ell_\mathrm{min},\ell_\mathrm{max}) = f_\mathrm{sky}\sum_{\ell = \ell_\mathrm{min}}^{\ell_\mathrm{max}}\frac{2\ell+1}{2}\mathrm{tr}\left( \Delta \boldsymbol{C}_\ell \boldsymbol{C}^{-1}_\ell\Delta \boldsymbol{C}_\ell \boldsymbol{C}^{-1}_\ell\right)\;,
\end{equation}
note that $\boldsymbol{C}_\ell$ is the matrix with the components $C^{(ij)}$. The factor $f_\mathrm{sky}$ takes into account the observed sky fraction. 
Using \Cref{eq:integrated_error} we rewrite the previous equation as a Riemann sum
\begin{equation}
\label{eq:delta_chi2_2}
\begin{split}
    \Delta\chi^2 (\ell_\mathrm{min},\ell_\mathrm{max}) = f_\mathrm{sky}& \sum_{\ell = \ell_\mathrm{min}}^{\ell_\mathrm{max}}\frac{2\ell+1}{2}\\ 
    &\times\sum_{r,s,i,j}\mathrm{tr}\left(\frac{\delta\boldsymbol{C}_\ell}{\delta_1 n^{(i)}(\chi_r)}\boldsymbol{C}^{-1}_\ell\frac{\delta\boldsymbol{C}_\ell}{\delta_1 n^{(j)}(\chi_s)} \boldsymbol{C}^{-1}_\ell\right)\\ & \times \mathcal{D}\chi_r\mathcal{D}\chi_s \Delta n^{(i)}(\chi_r)\Delta n^{(j)}(\chi_s)\;,
    \end{split}
\end{equation}
with the measure $\mathcal{D}\chi_r$.
If we define the Fisher matrix in this case as:
\begin{equation}
\label{eq:fisher_matrix}
    F_{\alpha\beta} = f_\mathrm{sky}    \sum_{\ell = \ell_\mathrm{min}}^{\ell_\mathrm{max}}\frac{2\ell+1}{2}\mathrm{tr}\left( \frac{\delta\boldsymbol{C}_\ell}{\delta_1 n_\alpha } \boldsymbol{C}^{-1}_\ell\frac{\delta\boldsymbol{C}_\ell}{\delta_1 n_\beta} \boldsymbol{C}^{-1}_\ell\right)\mathcal{D}\chi_{r(\alpha)}\mathcal{D}\chi_{s(\beta)}\;,
\end{equation}
where we labelled $n^{(i)}(\chi_r) \to n_{\alpha}$, we recover for a difference in $\chi^2$ using a scalar product on the finite-dimensional Hilbert space of shifts in the redshift distribution where the Fisher matrix acts as a norm-inducing metric
\begin{equation}
\label{eq:chi2}
    \Delta\chi^2 = F({\boldsymbol{\Delta n}},{\boldsymbol{\Delta n}})\equiv {\boldsymbol{\Delta n}}^T{\boldsymbol{F}}{\boldsymbol{\Delta n}}\;,
\end{equation}
where $\boldsymbol{\Delta n}$ is the vector containing shifts of the components $n_\alpha$. 

The Fisher matrix, \Cref{eq:fisher_matrix}, describes, how well the shifts $n_\alpha$ can be determined by a measurement of the angular power spectra $\boldsymbol{C}_\alpha$ given certain survey settings. If one would try to measure all possible perturbations, neighbouring $\delta n(\chi)$ are strongly correlated. This is, however not the question we would like to ask in this work. Instead, we want to look at the situation which we allow any perturbation $\boldsymbol{\Delta n}$, irrespective of the correlation. Therefore, by turning this argument around, we only use the diagonal part of the Fisher matrix. 

Lastly one should note that the functional derivative is strictly defined as a limiting process for infinitesimally small perturbation to the function at hand. The relation in general can be non-linear, but as long as relative perturbations to the function are small with respect to unity, these non-linear contributions are sub-dominant. Especially for surveys with tight requirements on the SRDs this is essentially always fulfilled.

\begin{figure}
    \centering
    \includegraphics[width = 0.45\textwidth]{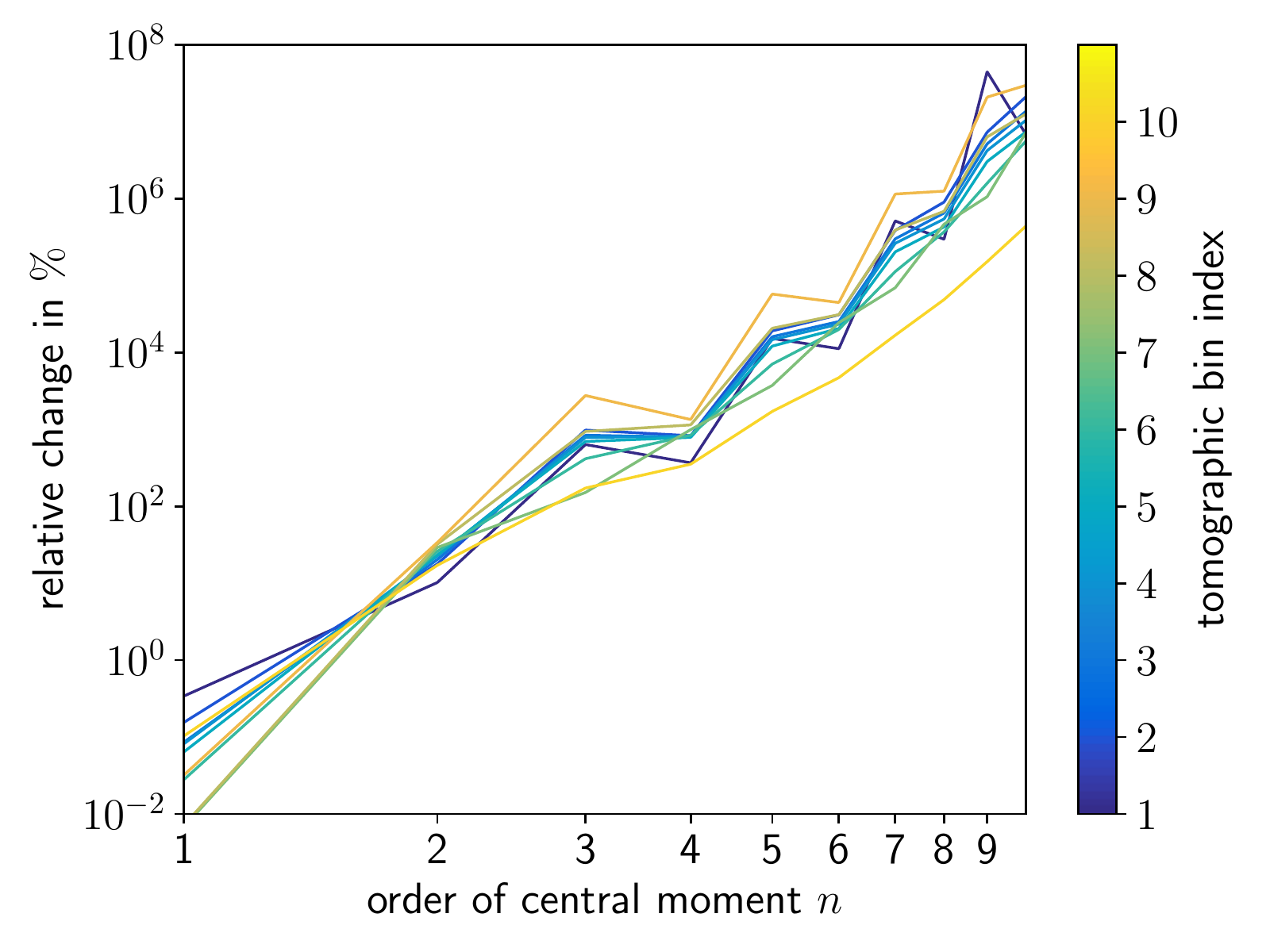}
    \caption{Allowed relative change in per-cent of the central moment of the SRD in each tomographic bin. The changes are calculated from the perturbed SRD distributions as shown in \Cref{fig:allowed_ecl}.}
    \label{fig:moments_euclid}
\end{figure}

\begin{figure}
    \centering
    \includegraphics[width = .45\textwidth]{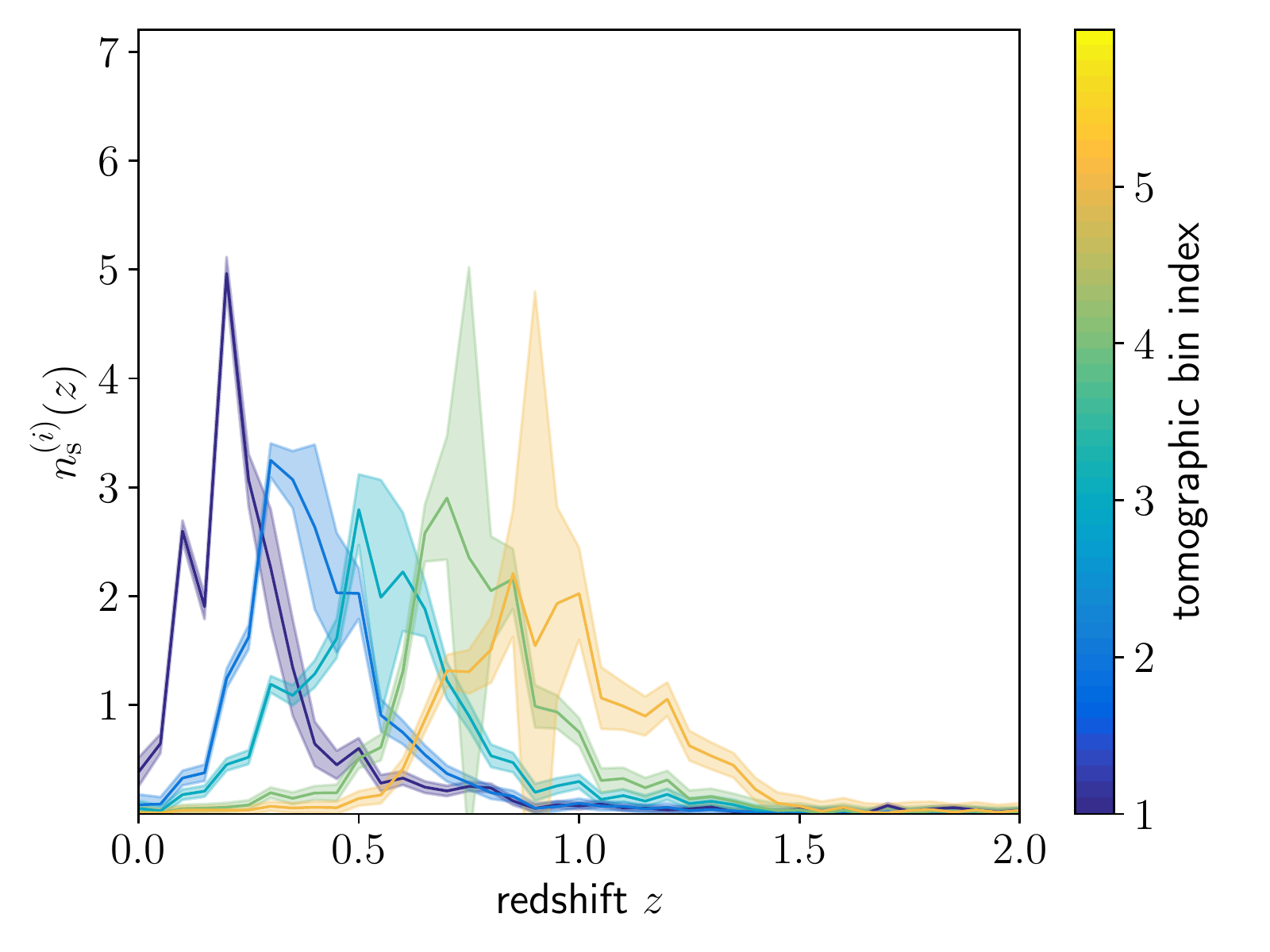}
    \caption{Allowed perturbation for KV450 to the SRD for the 5 tomographic source bins. Solid lines show the fiducial SRD, while the bands show the allowed perturbation to it, \revision{corresponding to a $\Delta\chi^2 = 1$.}}
    \label{fig:allowed_kv450}
\end{figure}

\section{Results}
\label{sec_results}
\subsection{Allowed Perturbations to the Source Redshift Distribution}
First, we will look at the allowed perturbations to the SRD by allowing for a total $\Delta\chi^2$ of unity, corresponding to a one $\sigma$ shift of a linear model parameter. Clearly, there are many different solutions $\boldsymbol{\Delta n}$ that satisfy $\Delta\chi^2 = 1$ subject to \Cref{eq:chi2}. To show the structure of the Fisher matrix we therefore distribute the allowed $\Delta\chi^2$ per $\Delta n_\alpha$ equally. 

We will assume EUCLID specifications for the survey as given in \citet{blanchard_euclid_2020} and assume $n_\mathrm{tomo}= 10$ tomographic bins, a sky fraction of $0.3$. 
Furthermore, we will collect multipoles between $\ell_\mathrm{min} = 10$ and $\ell_\mathrm{max} = 3000$. We then calculate the diagonal Fisher matrix from \Cref{eq:fisher_matrix} and distribute the errors equally as described above. This results in a possible realisation of $\boldsymbol{\Delta n}$ yielding $\Delta\chi^2 = 1$ subject to the constraint \Cref{eq:srd_norm}. 
\Cref{fig:allowed_ecl} shows the resulting perturbed SRDs. The solid lines show the fiducial SRD, while the shaded areas show the allowed perturbations to not cause a bias of more than 1 $\sigma$ for a linear model parameter. Lastly, the tomographic bin index is shown as a colour bar. The general trend is very clear, the allowed perturbations become very large around a small interval $\Delta\chi$ around the mean of the distributions. For most tomographic bins this coincides with the peak of the distribution as they are very close to Gaussian. Only for the first and the last bin, these spikes are a bit offset since the distributions are a bit more asymmetric. This already confirms that the most important part of the SRDs in cosmic shear measurements is to calibrate the mean redshift of each tomographic bin very well. Furthermore, we observe that the spikes tend to be narrower at higher redshifts, indicating that the uncertainty on the mean of the SRD is more important at higher redshifts \revision{(see also \cref{fig:moments_euclid}).}

{\revision{We want to stress again, that this is just one realization of $\boldsymbol{\Delta n}$ that produces a $\Delta\chi^2 = 1$, but by distributing the errors equally, it is possible to see, which perturbations the final measurement is most sensitive to. However, the uncertainties should not be \revision{taken at face value} and are extreme values, they just give a general trend. Furthermore, it is also important to notice that we treat the variation at each co-moving distance (or redshift) as independent. While the exact spacing does not affect the results as long as the Riemann sum, \Cref{eq:delta_chi2_2}, converges, an issue can arise, however, when inverting the Fisher matrix, \Cref{eq:fisher_matrix}, for finely sampled data points as it can become degenerate. This issue was discussed as well in \citet{2023arXiv230617224K} and requires the definition of an equivalence class by restricting possible variations of $\Delta n $ to the quotient space of those functions (SRDs) which the data can distinguish. Having said this, \Cref{fig:allowed_ecl}, should be understood as a sensitivity scan of the angular power spectra to the SRD only.}}

Next, we use the perturbed SRDs to calculate their central moments $\mu_n$:
\begin{equation}
    \mu_n \coloneqq E[(X-E[X])^n] = \int p(x)(x-\mu)^n\mathrm{d}x\;,
\end{equation}
for a probability distribution function $p(x)$ with mean $\mu$. The perturbed SRDs are used to calculate the change in the central moments relative to the fiducial SRD. \Cref{fig:moments_euclid} shows the resulting relative change for all tomographic bins as a function of the order of the central moment. Clearly, the first moment is most important and while the second one still needs to be known at a 10$\%$ level, all higher-order moments are essentially unimportant. This is of course reminiscent of the behaviour observed in \Cref{fig:allowed_ecl}, where the perturbations are such that they essentially fix the mean. It is of course entirely possible, that we alter the shape of the distribution differently, but still achieve the desired accuracy. {\revision{As mentioned before, there are perturbations to the SRD which can source larger changes in the moments, what we would like to show here is the relative importance of the moments. How an ensemble of perturbations to the SRD will affect an actual cosmological analysis is studied in the next section, where the correlations at different redshifts are taken into account. }}

Nonetheless, the results show that for the SRD for cosmic shear, only the mean redshift and the width are important with the former influencing the result way more (by over an order of magnitude). In \Cref{sec:m_and_v_kv450} we sample from the allowed changes in the SRD and show the relative difference of the first two moments to illustrate their scatter.

\begin{figure}
    \centering
    \includegraphics[width = 0.45\textwidth]{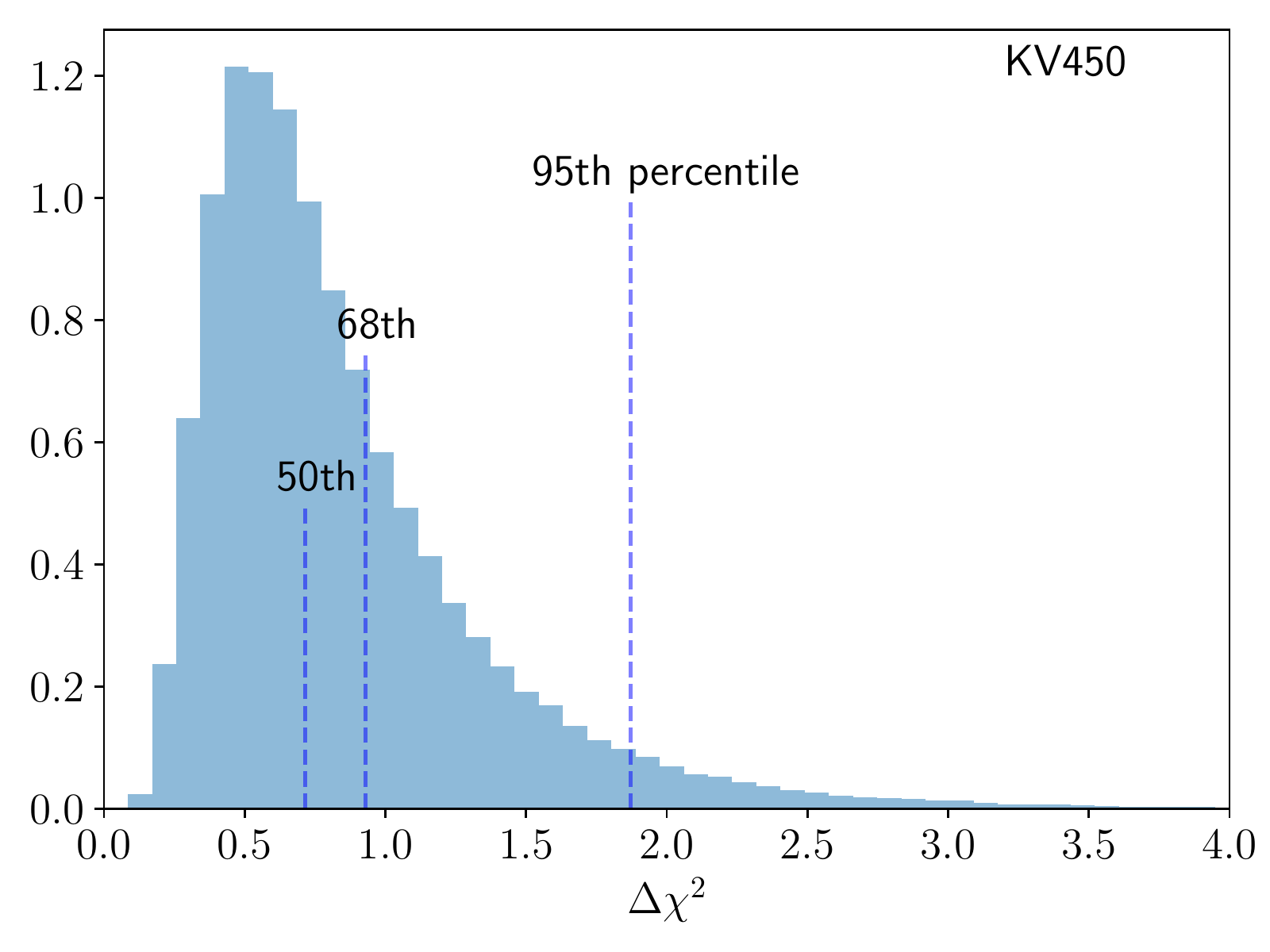}
    \caption{$\Delta\chi^2$ for $10^6$ realisations of $\boldsymbol{\Delta n}$ from the  $\boldsymbol{C}^\mathrm{KV450}_{n(\chi)}$. We also show the 50, 68 and 95 percentiles. 
    }
    \label{fig:dchi2_kv450}
\end{figure}

\subsection{Propagating Redshift Errors}
In this section, we will revisit the KV450 data for the SRD \citep{hildebrandt_kidsviking-450_2020}. This data set is used since it includes a covariance matrix from the direct calibration (DIR), for the clustering redshifts \citep{busch_testing_2020} or the self-organising maps \citep{wright_photometric_2020} no bootstrap covariance was estimated so far.

For completeness, the allowed perturbations are shown in \Cref{fig:allowed_kv450}. Due to the lower signal-to-noise ratio of the measurement, the allowed perturbations are much larger than in the previous case. The features, however, are very similar. 

Since we are expressing everything in co-moving distance, the covariance matrix needs to be transformed accordingly. Let $\boldsymbol{C}^\mathrm{KV450}_{n(z)}$ be the covariance matrix in $n(z)$ space, the transformed covariance is then 
\begin{equation}
    \boldsymbol{C}^\mathrm{KV450}_{n(\chi)} = \boldsymbol{J}^T\boldsymbol{C}^\mathrm{KV450}_{n(z)}\boldsymbol{J}\;,
\end{equation}
where $\boldsymbol{J}$ is the Jacobian with components $J^{i}_{\;j} = \delta^{i}_j\mathrm{d}z/\mathrm{d}\chi$. Alternatively, the Fisher matrix of the SRD perturbations can be expressed in redshift space by the inverse transform.

Perturbations $\boldsymbol{\Delta n}$ are now sampled from $\boldsymbol{C}^\mathrm{KV450}_{n(\chi)}$ and propagated to obtain $\Delta\chi^2$ according to \Cref{eq:delta_chi2}. If the redshift errors as given in $\boldsymbol{C}^\mathrm{KV450}_{n(\chi)}$ are sufficiently small to not produce a significant bias in the cosmological parameters such as $S_8$ we expect most realisations \citep[i.e. 68$\%$][]{hildebrandt_kidsviking-450_2020} to yield $\Delta\chi^2 < 1$. \Cref{fig:dchi2_kv450} shows the resulting distribution in $\Delta\chi^2$ for the $10^6$ realizations of $\boldsymbol{\Delta n}$ for KV450. The vertical dashed lines show the 50th, 68th and 95th percentile. It is clear from this plot that the precision of the SRD used in KV450 is high enough to not yield any spurious detection in the final parameter constraints since the 68th percentile is still well below unity. \revision{This is also in agreement with Figure 6 in \citet{hildebrandt_kidsviking-450_2020}, showing almost identical $S_8$ results when ignoring the uncertainties in the SRD (the shift parameters in this case).} 

{\revision{Before moving on, two comments about the limitations of the method are in order:}}

{\revision{
$(i)$ The produced realisations from the KV-450 covariance matrix produce perturbations $\Delta n(z)$ which are not necessarily small. Therefore the functional derivative is just an first-order approximation to the non-linear dependency on the SRD fluctuation. However, as surveys become more constraining the requirements on the SRD precision become tighter as well, making the method more applicable in the future. Furthermore, using just a linear model is a fair comparison to what has been done in KV-450, where also Gaussian error propagation of the mean shifts was used to estimate the induced bias on cosmological parameters.
}}

{\revision{
$(ii)$ As mentioned earlier we collect multipoles from 10 to 3000 for the signal of EUCLID. KV-450 on the other hand only measures real space correlation functions over a finite angular range. In principle correlation functions in configuration space get contributions from all multipoles at every angular scale (see \cref{sec:observables}). Here we choose $\ell_\mathrm{min}$ and $\ell_\mathrm{max}$ such that they match the inverse angular scales used for the KV-450, $[0.5,300]\;\mathrm{arcmin}$, effectively amounting to the multipole range used for the KiDS bandpower analysis $\ell\in[100,1500]$ \citep{joachimi_kids-1000_2021}. In general, the choice of the scales involved in the analysis will change the requirement on SRD uncertainties slightly since different scales obtain most of their signal from different redshifts.}}
\begin{figure}
    \centering
    \includegraphics[width=.45\textwidth]{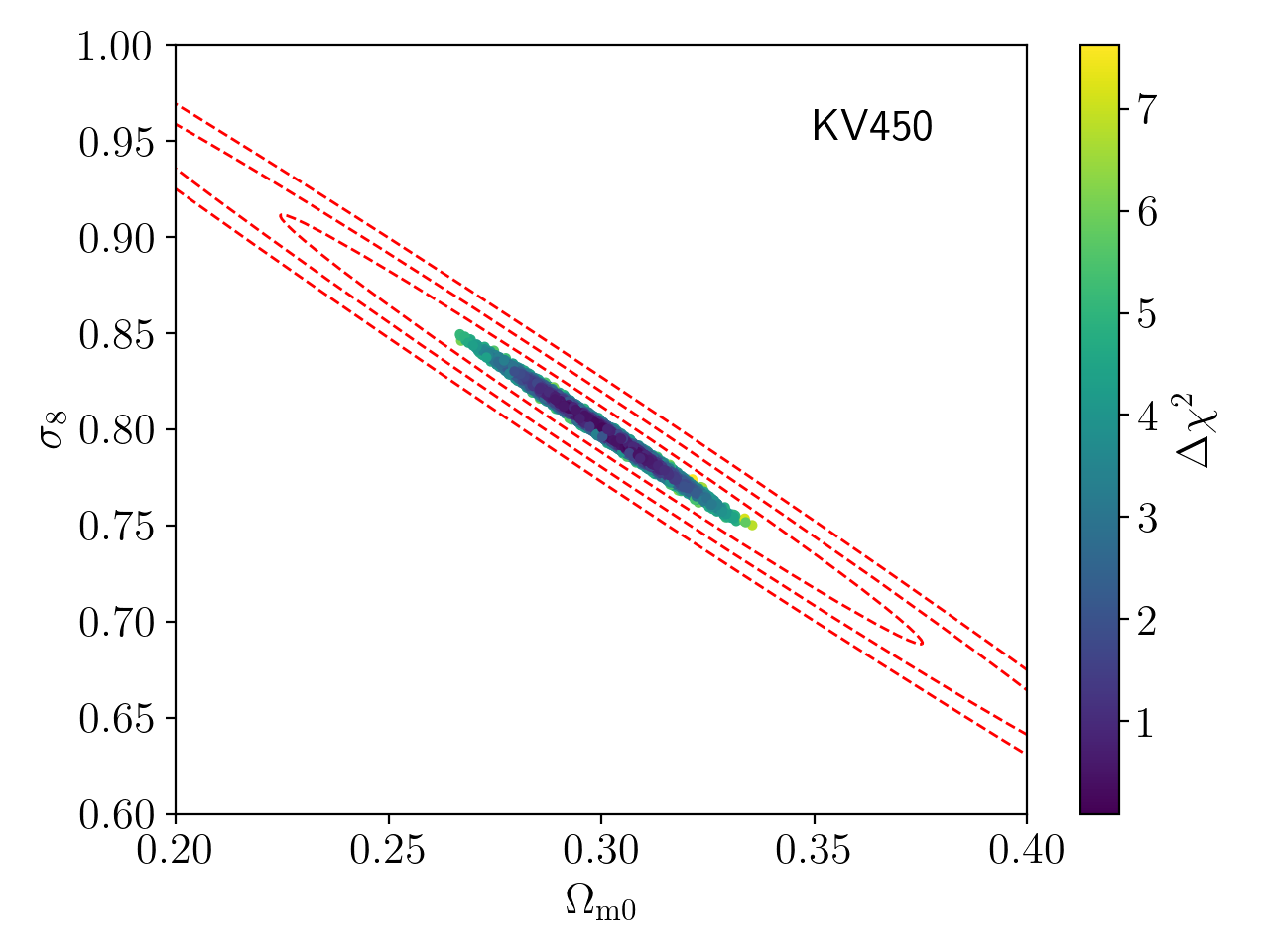}
    \caption{\revision{The scatter points show the induced shifts by the photo-$z$ uncertainty in the $\Omega_\mathrm{m0}$-$\sigma_8$-plane, derived from the $\Delta\chi^2$ (colour-coded) of \Cref{fig:dchi2_kv450}. In red we show the contour from the Fisher matrix for KV450 enclosing the 1$\sigma$ confidence interval.}}
    \label{fig:kv_450_bias}
\end{figure}

\begin{figure}
    \centering
    \includegraphics[width = .45\textwidth]{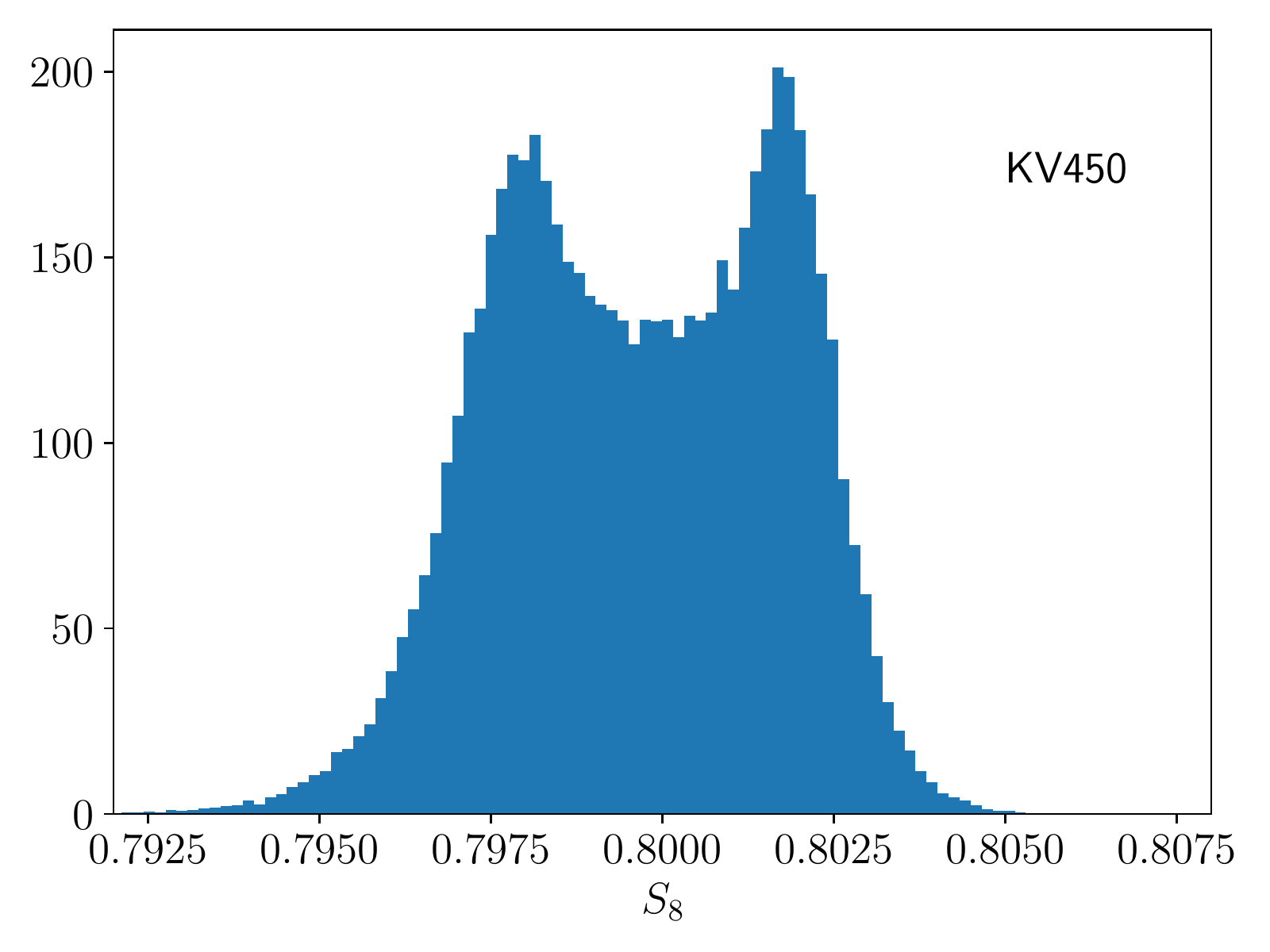}
    \caption{Induced scatter on the $S_8 = \sigma_8\sqrt{\Omega_\mathrm{m0}/0.3}$ parameter. This is directly derived from the samples of \Cref{fig:kv_450_bias}. The scatter is roughly 15 per-cent of the statistical error budget reported in \citet{Hildebrandt:2016iqg,hildebrandt_kidsviking-450_2020}.}
    \label{fig:s8}
\end{figure}

One could now further propagate these uncertainties into cosmological parameters using the corresponding Fisher matrix. For a given shift in the SRD $\boldsymbol{\Delta n}$, the corresponding shifts in the cosmological parameters, $\boldsymbol{\Delta\theta}$ can be calculated:
\begin{equation}
    \Delta\theta^{i} = - (F^{-1})^{i}_{\;k}\mathcal{F}^{k}_{\;\beta}\Delta n^\beta\;,
\end{equation}
where Greek indices run over the perturbations in the SRD, while Latin indices label cosmological parameters. Here we assumed the sum convention. $\mathcal{F}^{i}_{\;\alpha}$ hence is the mixed pseudo Fisher matrix:
\begin{equation}
    \mathcal{F}^{i}_{\;\alpha} = -E\left[\frac{\partial\ln L}{\partial\theta_i }\frac{\delta\ln L}{\delta n^\alpha }\mathcal{D}\chi_{r(\alpha)}\right]
\end{equation}
and its inverse is a pseudo inverse. Since the inversion of this matrix is not necessarily stable we choose to go another route here. Since the distribution of $\Delta\chi^2$ is known, we are interested in samples of cosmological parameters with the same $\Delta\chi^2$ with respect to the best-fit value. For a Gaussian posterior in one dimension this would amount to a distribution such that the absolute value of each sample is fixed to $\sqrt{\Delta\theta^2}$. We sample from a standard Gaussian distribution and modify its width by $\sqrt{\Delta\theta^2}$. This Gaussian is then mapped into the frame of the cosmological parameters under consideration via the Cholesky decomposition of the Fisher matrix of the cosmological parameters. In \Cref{fig:kv_450_bias} we apply this procedure to the $\Delta\chi^2$ distribution of KV450 (\Cref{fig:dchi2_kv450}). Each dot represents one sample of the $\Delta\chi^2$ distribution with its value shown as a colour bar. It can be seen as the geodesic distance to the fiducial value for the cosmological parameters in the parameter manifold \citep{giesel_information_2021}. The red contours depict the expected $1,2,3\sigma$ confidence regions from the Fisher forecast for KV450. Since in the original analysis more than the two parameters here were used, we re-scale the $\Delta\chi^2$ accordingly, in particular by the $\chi^2$ quantile function $\chi^2_k(p)$, where $k = 10$ is the number of parameters in the actual analysis \citep{Hildebrandt:2016iqg} and $p = 0.68$. This is done to obtain a fair comparison.
It is clear from the plot, that all samples for the photometric redshift distribution lie well within the $1\sigma$ contour. Furthermore, it should be noted that we are considering a very idealised forecast with two free parameters and no systematics here. The procedure, however, can be generalized to any number of parameters. Furthermore, one can apply the same analysis to a full Monte-Carlo-Markov-Chain (MCMC) by matching those samples which are $\Delta\chi^2$ away from the maximum likelihood of the MCMC. 
Lastly, the samples from \Cref{fig:kv_450_bias} can be mapped to $S_8 = \sigma_8\sqrt{\Omega_\mathrm{m0}/0.3}$. \Cref{fig:s8} shows the resulting histogram of the scatter due to the photo-$z$ uncertainties. Comparing this to $\Delta S_8 = 0.076$ at $68\%$ confidence \citep{hildebrandt_kidsviking-450_2020} shows that the scatter induced by the redshift uncertainties as sampled from the KV450 SRD covariance have a small effect on the overall error budget. In \citet{Hildebrandt:2016iqg} a Fisher matrix method for the shifts of the mean of the SRDs was investigated as a source of systematics, which found similar results to the ones presented here. The main difference between the two methods is that we allow for general perturbations to the redshift distribution (provided their correlation is given). Generalizing the procedure in \citet{Hildebrandt:2016iqg} to moments higher than the variance is bound to fail (see \Cref{ssec:edgeworth expansion}). However, we would also conclude that even for EUCLID, the analysis of the first two moments is probably sufficient. 

In \cref{sec:m_and_v_kv450} the mean and standard deviation of each SRD in the five tomographic bins are shown for the realisations used in this section as sampled from the DIR covariance matrix. \Cref{fig:momentskv450} shows a very similar behaviour to what we found in \Cref{fig:moments_euclid}. In particular \revision{we find that} the mean scatters less at higher redshifts, while the standard deviation scatters roughly equally for most of the bins.

\revision{We close the section with a general discussion about the usage of $\Delta\chi^2$ or directly assessing uncertainties in cosmological parameters.} It is in general advantageous to make accuracy assessments for the SRD using the $\Delta\chi^2$ and not by inverting the Fisher matrix for the parameters of interest to obtain the shift values for those. The reason for this is that $\Delta\chi^2$ is an invariant quantity, while shifts in parameter space are dependent on the specific model choice. The only caveat in the $\Delta\chi^2$ is that the number of parameters must be taken into account, this is, however, much easier than calculating the Fisher matrix.

\section{Conclusions}
\label{sec:conclusions}
In this paper, we have analysed the dependence of the cosmic shear angular power spectrum on the SRD. This has been done by employing functional derivatives of the cosmic shear $C_\ell$ with respect to the SRD at a fixed co-moving distance $\chi_0$. By integrating over the introduced error we estimated the $\Delta\chi^2$ introduced by arbitrary uncertainties in the SRD. We applied our method to a cosmic shear survey with EUCLID specifications and KV450 since a covariance of the SRD estimate was given. Our main findings can be summarised as follows:

\begin{enumerate}
    \item Allowed perturbations of the SRD are such that they preserve the mean of the underlying distribution. If they do, they can be rather larger, even for a survey like EUCLID. This is in line with the common practice of using only shifted means of the underlying redshift distribution. 
    \item To achieve the accuracy required for EUCLID, the mean of the redshift distribution needs to be determined between 1 and 0.01 per-cent, depending on the tomographic bin under consideration. The variance of the SRD is still important at the 10 per-cent level. There is still some sensitivity left in the skewness, but all other moments are not relevant. 
    \item We performed a simplistic analysis of the KV450 SRDs to check whether they fulfil the requirements and found that the uncertainties, in this very idealised scenario, only yield biases up to $1\sigma$ in the final constraints. In a full analysis, this bias would be even smaller. Thus confirming the redshift calibration used in KV450.
        \item Even for EUCLID it is most likely not necessary to investigate moments of the redshift distribution $n>2$. This conclusion could change for different settings and self-calibration methods.
    \item The procedure outlined here has the advantage of being very cheap computationally since the functional derivatives only need to be computed once. It is then only a matter of sampling from the underlying SRD and propagating these perturbations with the previously calculated functional derivative. It is hence not necessary to push thousands of realisations of the SRD through the analysis pipeline.
\end{enumerate}
The method outlined here can thus be used to analyse whether a perturbation in the SRD still fulfils the requirements of a given experiment so that no biases of model parameters are introduced. It allows for arbitrary perturbations to the SRD without requiring a fit to the actual distribution. We intend to apply the presented method to the updated SRDs of KiDS in the future.

For the interested reader the appendices \Cref{ssec:edgeworth expansion} - \Cref{sec:nonlimber} discuss various aspects of the analysis which could be refined in future work. In particular, we look at the Edgeworth expansion of the SRD in \Cref{ssec:edgeworth expansion}, i.e. an expansion in the cumulants of the underlying SRDs. However, we find that, even for a realistic setting, the Edgeworth expansion cannot reproduce the original SRDs if cumulants $n>2$ are considered.

\section*{Data Availability} The data underlying this article will be shared on reasonable request to the corresponding author.

\section*{Acknowledgments}
RR would like to thank Hendrik Hildebrandt and Bj\"orn Malte Sch\"afer for insightful discussions and comments on the manuscript.
RR is supported by the European Research Council (Grant No. 770935). 
\revision{RR would like to thank an anonymous referee for very valuable comments regarding the presentation of the manuscript.}
\bibliographystyle{mnras}
\bibliography{MyLibrary}

\appendix

\section{Edgeworth expansion}
\label{ssec:edgeworth expansion}
\begin{figure}
    \centering
    \includegraphics[width = .45\textwidth]{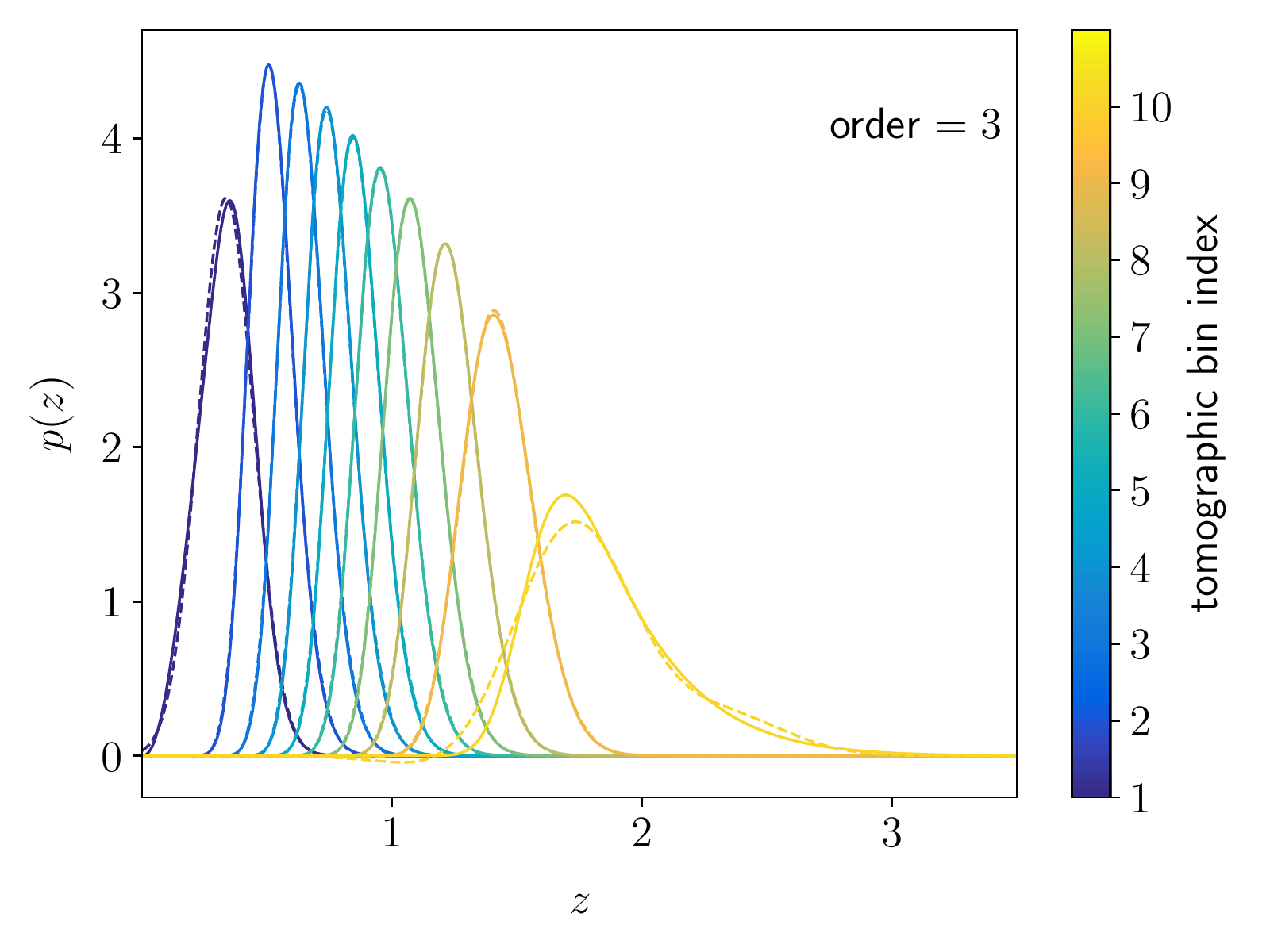}
        \includegraphics[width = .45\textwidth]{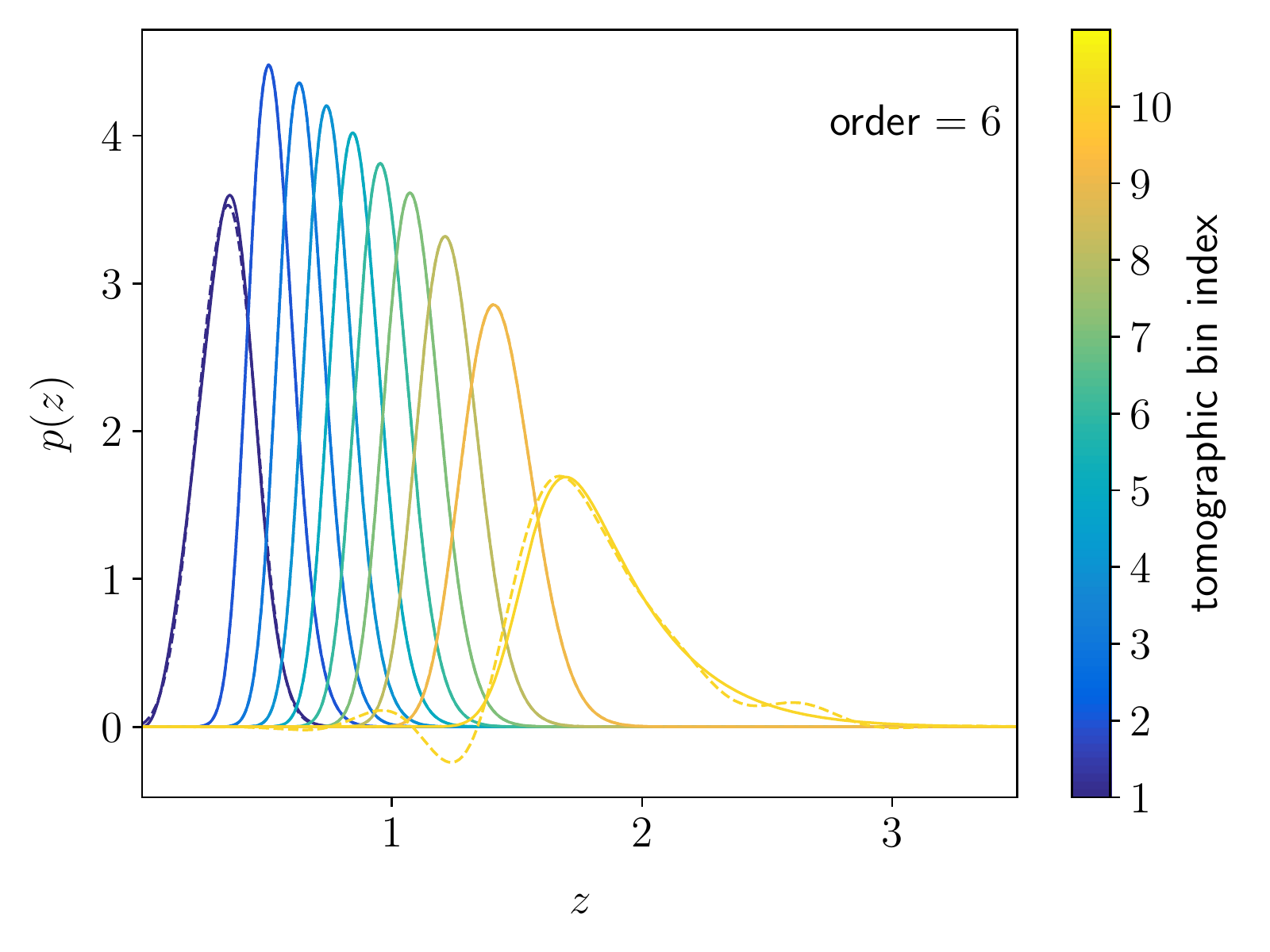}
    \caption{SRD for EUCLID in all 10 tomographic bins. Solid lines represent the fiducial SRD, while dashed lines represent their respective Edgeworth expansion. Cumulants up to order $n=3\;,6$ are used respectively.}
    \label{fig:srd_edgeworth}
\end{figure}
In this section, we employ an Edgeworth expansion for the photo-$z$ distribution. The Edgeworth expansion is an asymptotic expansion (in contrast to the Gram-Charlier expansion). Starting from the characteristic function \citep{blinnikov_expansions_1998}.
\begin{equation}
    \varphi^{(j)}_Z(t) = \mathrm{E}_{n^{(j)}(z)}\left[\mathrm{e}^{\mathrm{i}tZ}\right]\;,
\end{equation}
i.e. the Fourier transform of the probability density $n^{(j)}(z)$. With the definition of the moments $\Tilde{\mu}_n$, the Taylor expansion of the characteristic function is 
\begin{equation}
     \varphi^{(j)}_Z(t) = 1 + \sum_{n_1}^\infty \frac{\Tilde{\mu}_n}{n!}(\mathrm{i}t)^n\;.
\end{equation}
\begin{figure*}
    \centering
    \includegraphics[width =.9\textwidth]{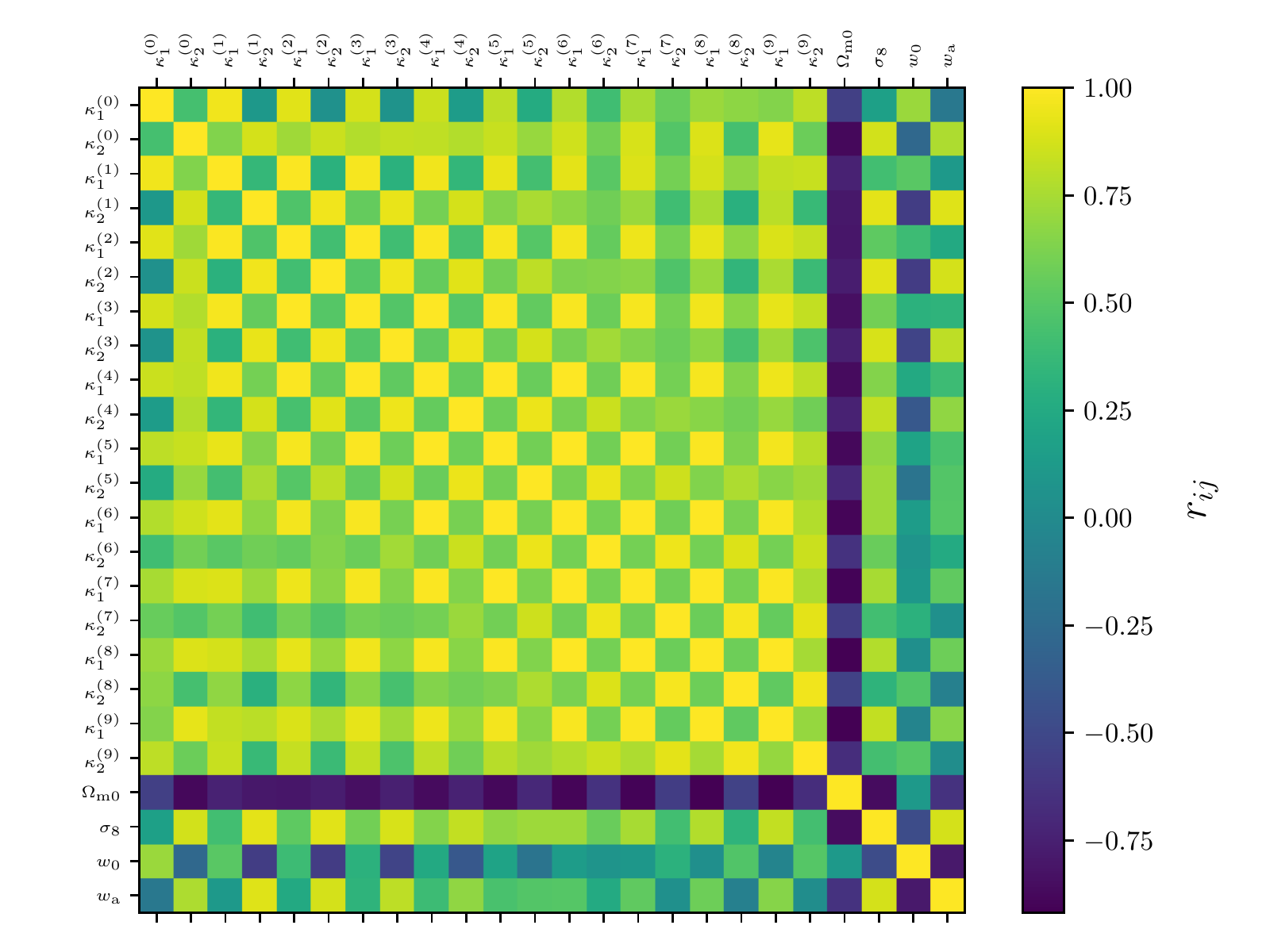}
    \caption{Pearson correlation coefficient for the joint covariance matrix of the first two cumulants of the EUCLID like survey and four cosmological parameters.}
    \label{fig:pearson}
\end{figure*}
The logarithm of the characteristic function is the cumulant, $\kappa_n$, generating function
\begin{equation}
    \kappa_n =\frac{1}{\mathrm{i}^n}\frac{\mathrm{d}^n}{\mathrm{d}t^n}\log\varphi^{(j)}_Z(t)\,\bigg|_{t=0}\;.
\end{equation}
Using this definition one can relate the cumulants to the moments 
\begin{equation}
    \kappa_n = n!\sum_{\{k_m \}}(-1)^{r-1}(r-1)!\prod_{m=1}^n\frac{1}{k_m!}\left(\frac{\Tilde{\mu}_m}{m!}\right)^{k_m}\;,
\end{equation}
where $\{k_m\}$ denotes the set of all solutions to the Diophantine equation 
\begin{equation}
    \sum_{a=1}^n ak_a - n = 0\;.
\end{equation}
If a distribution is then expanded as an asymptotic series around a normal distribution one finds
\begin{equation}
\begin{split}
    n(z) = & \ \frac{1}{\sqrt{2\pi\kappa_2}}\exp{\left(-\frac{(z-\kappa_1)^2}{2\kappa_2}\right)} \\ 
    & \times\Bigg[1+\sum_{s=1}^\infty\kappa^{s/2}_2\sum_{\{k_m\}}\mathrm{He}_{s+2r}\left(\frac{z}{{\kappa^{1/2}_2}}\right)\prod_{m=1}^s\frac{1}{k_m!}\left(\frac{\lambda_{m+2}}{(m+2)!}\right)^{k_m}\Bigg]\\ 
    \equiv & \ n_G(z)(1+\mathrm{Eg}(z))  ,
    \end{split}
\end{equation}
where $\lambda_n \coloneqq \kappa_n/\kappa^{n/2}_2$. We are now interested in the sensitivity of the distribution with respect to its cumulants. Here the cases $n=1,2$ are a bit special:
\begin{equation}
    \frac{\partial n(z)}{\partial \kappa_1} = n(z)\frac{z-\kappa_1}{\kappa_2}
\end{equation}
and for $\kappa_2$
\begin{equation}
    \frac{\partial n(z)}{\partial \kappa_2}=\frac{1}{2\kappa^{1/2}_2}\left[n(z)\left( \frac{(z-\kappa_1)^2}{\kappa_2}-1\right) + n_{G}(z)\frac{\partial \mathrm{Eg}(z)}{\partial\kappa^{1/2}_2}\right]\;,
\end{equation}
where
\begin{equation}
\begin{split}
 &       \frac{\partial \mathrm{Eg}(z)}{\partial\kappa^{1/2}_2} =  \sum_{s=1}^\infty\sum_{\{k_m\}}\kappa^{s/2}_2\mathcal{P}(s,\{k_m\})\Bigg[\mathrm{He}_{2r+s}\left(\frac{z}{{\kappa^{1/2}_2}}\right) \\ & \times\left(\frac{s}{\kappa^{1/2}_2}-\sum_{a}^s \frac{k_a(2a+2)}{\kappa^{k_a(a+1)+1/2}_2} \right) -(2r+s)\mathrm{He}_{2r+s-1}\left(\frac{z}{{\kappa^{3/2}_2}}\right)\frac{z}{{\kappa_2}}\Bigg]\;,
        \end{split}
\end{equation}
where we also defined the product:
\begin{equation}
    \mathcal{P}(s, \{k_m\}) \coloneqq \prod_{m=1}^s\frac{1}{k_m!}\left(\frac{\kappa_{m+2}}{(m+2)!\kappa^{2m+2}_2}\right)^{k_m}\;.
\end{equation}
For all cumulants with $n\geq 3$ one finds:
\begin{equation}
    \frac{\partial n(z)}{\partial \kappa_n} = n_G(z)\sum_{s=1}^\infty\sum_{\{k_m\}}\kappa^{s/2}_2\mathcal{P}(s,\{k_m\}) \mathrm{He}_{2r+s}\left(\frac{z}{{\kappa^{1/2}_2}}\right)\frac{k_{n-2}}{\kappa_n}\;.
\end{equation}
It should be noted, however, that the Edgeworth expansion is not a convergent series but rather an asymptotic expansion. One therefore needs to check whether the expansion is a good approximation of the underlying distribution.

In this case, one can define the ordinary Fisher matrix using partial derivatives:
\begin{equation}
    F_{\kappa^{(i)}_m\kappa^{(j)}_n} = f_\mathrm{sky}    \sum_{\ell = \ell_\mathrm{min}}^{\ell_\mathrm{max}}\frac{2\ell+1}{2}\mathrm{tr}\left( \frac{\partial\boldsymbol{C}_\ell}{\partial \kappa^{(i)}_m } \boldsymbol{C}^{-1}_\ell\frac{\partial\boldsymbol{C}_\ell}{\partial\kappa^{(j)}_n} \boldsymbol{C}^{-1}_\ell\right) \;,
\end{equation}
where $\kappa^{(i)}_m$ is the $m$-th cumulant of the source-redshift distribution in the $i$-th tomographic bin. 

\Cref{fig:srd_edgeworth} shows the fiducial redshift distributions for EUCLID and their Edgeworth expanded approximations as solid and dashed lines respectively. The top plot uses the expansion up to  $\kappa_3$, while the bottom plot sums contributions up to $\kappa_6$. For all but the first and last tomographic bin, the Edgeworth series is a good approximation. This is expected as they are essentially Gaussian and therefore $\kappa_n \approx 0$ for $n > 2$. The first tomographic bin experiences boundary effects at $z=0$ and is therefore slightly skewed. This effect is even larger for the last tomographic bin, which has a very long tail to high redshifts. While the first bin can still be described 
by the Edgeworth expansion and the series converges, the 10th bin shows negative probability in the Edgeworth series already at third order. The situation becomes worse if higher-order cumulants are included.
This goes to show that even for such an idealized case as the EUCLID forecast, the use of the Edgeworth expansion can be very dangerous. 

For the case $n=2$ we show the Pearson correlation coefficient of the joint covariance matrix between the first three cumulants in each tomographic bin and four cosmological parameters in \Cref{fig:pearson}. We observe some correlations between the first and second moments of each tomographic bin. There is a very strong correlation between the first and second moment of two different redshift bins. Furthermore, one can see that parameters controlling the amplitude of the lensing spectrum are anti-correlated with the mean. We want to stress again, however, that the expansion, even in this case, is not convergent and results obtained with $n>2$ have thus to be taken with care.

\section{Photometric Galaxy Clustering}
\label{sec:photometric_clustering}
\begin{figure}
    \centering
    \includegraphics[width = .45\textwidth]{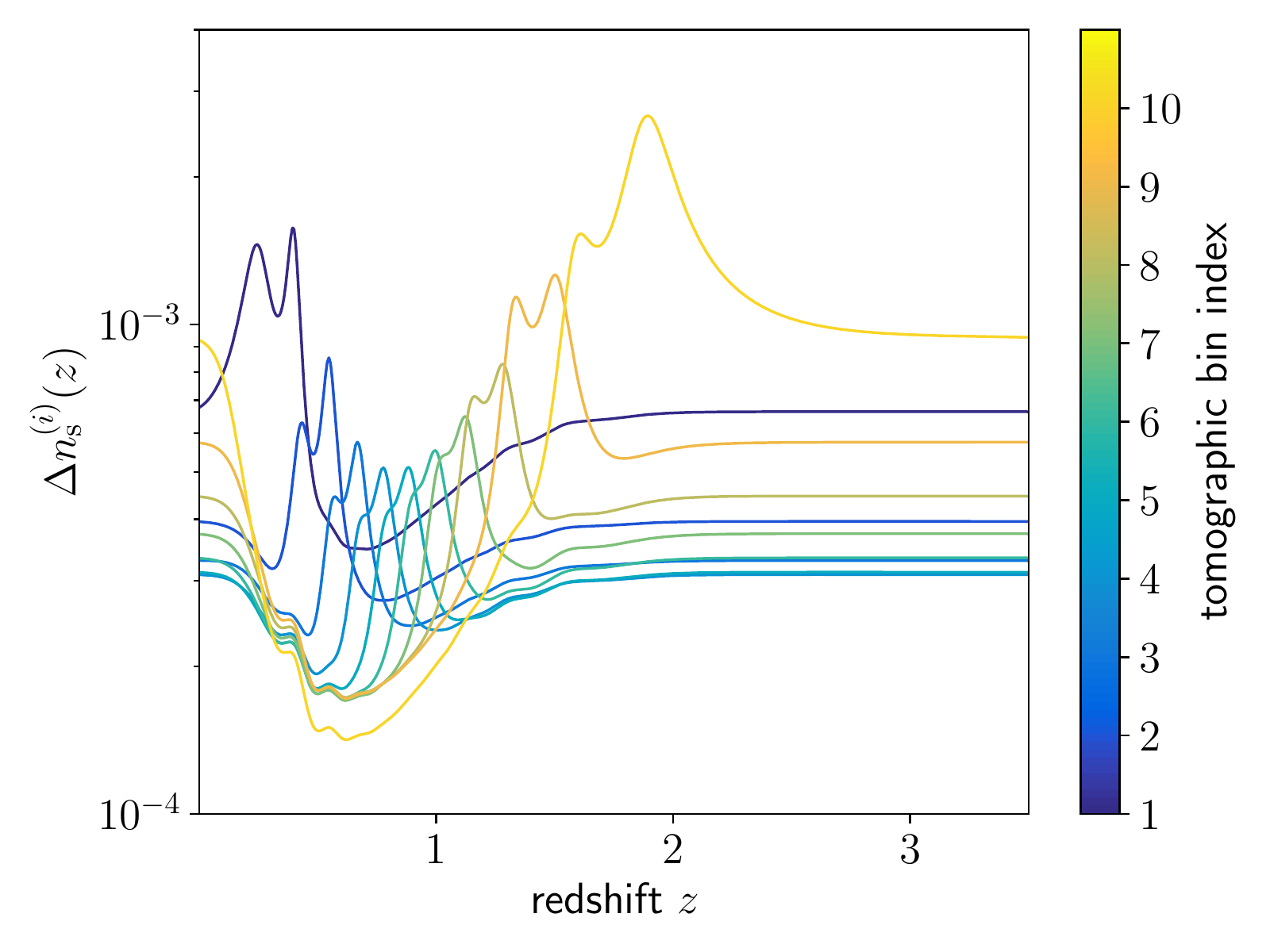}
    \caption{Allowed perturbations for the redshift distribution for photometric galaxy clustering.}
    \label{fig:allowed_gg}
\end{figure}

For photometric galaxy clustering, the procedure can be simply adopted by changing the weight function (up to galaxy bias, which we absorb in the power spectrum). Again by using the Limber projection:
\begin{equation}
    C^{g_ig_j}_\ell = \int_0^{\chi_\mathrm{H}}\frac{\mathrm{d}\chi}{\chi^2} W^{(i)}_g(\chi)W^{(j)}_g(\chi) P_{gg}\left(\frac{\ell + 0.5}{\chi},\chi\right)\;,
\end{equation}
with the galaxy power spectrum $P_{gg}$ and corresponding weights given by:
\begin{equation}
    W^{(i)}_g(\chi) = n^{(i)}_g(\chi)\;,
\end{equation}
therefore the functional derivative takes the very simple form
\begin{equation}
    \frac{\delta C^{g_ig_j}_\ell}{\delta n^a(\chi_0)} = \frac{P_{gg}\left(\frac{\ell + 0.5}{\chi},\chi\right)}{\chi^2} \left(n^{(j)}(x)\delta^\mathrm{K}_{ia} + n^{(i)}(x)\delta^\mathrm{K}_{ja}\right)\;.  
\end{equation}
We show the corresponding allowed perturbations (cf. \Cref{fig:allowed_ecl}) in \Cref{fig:allowed_gg} where the different shape compared to the cosmic shear case is clearly visible. While the latter had a distinctive peak around the mean of the distributions, the allowed perturbations of galaxy clustering have two maxima, indicating that also the width of the distribution is important. Overall the allowed perturbations are much smaller, this is, however, because the clustering signal is much larger in general.

\section{Distribution of the Mean and Variance}
\label{sec:m_and_v_kv450}
We show the relative difference between the mean redshift and the standard deviation of the SRD for each tomographic bin. As before we distinguish between the EUCLID's survey settings and KV450. In particular, we sample from the diagonal covariance obtained from the functional Fisher matrix as described in \Cref{sec_results} for the former, while we use the DIR covariance for the latter. 

\begin{figure*}
    \centering
    \includegraphics[width=0.45\textwidth]{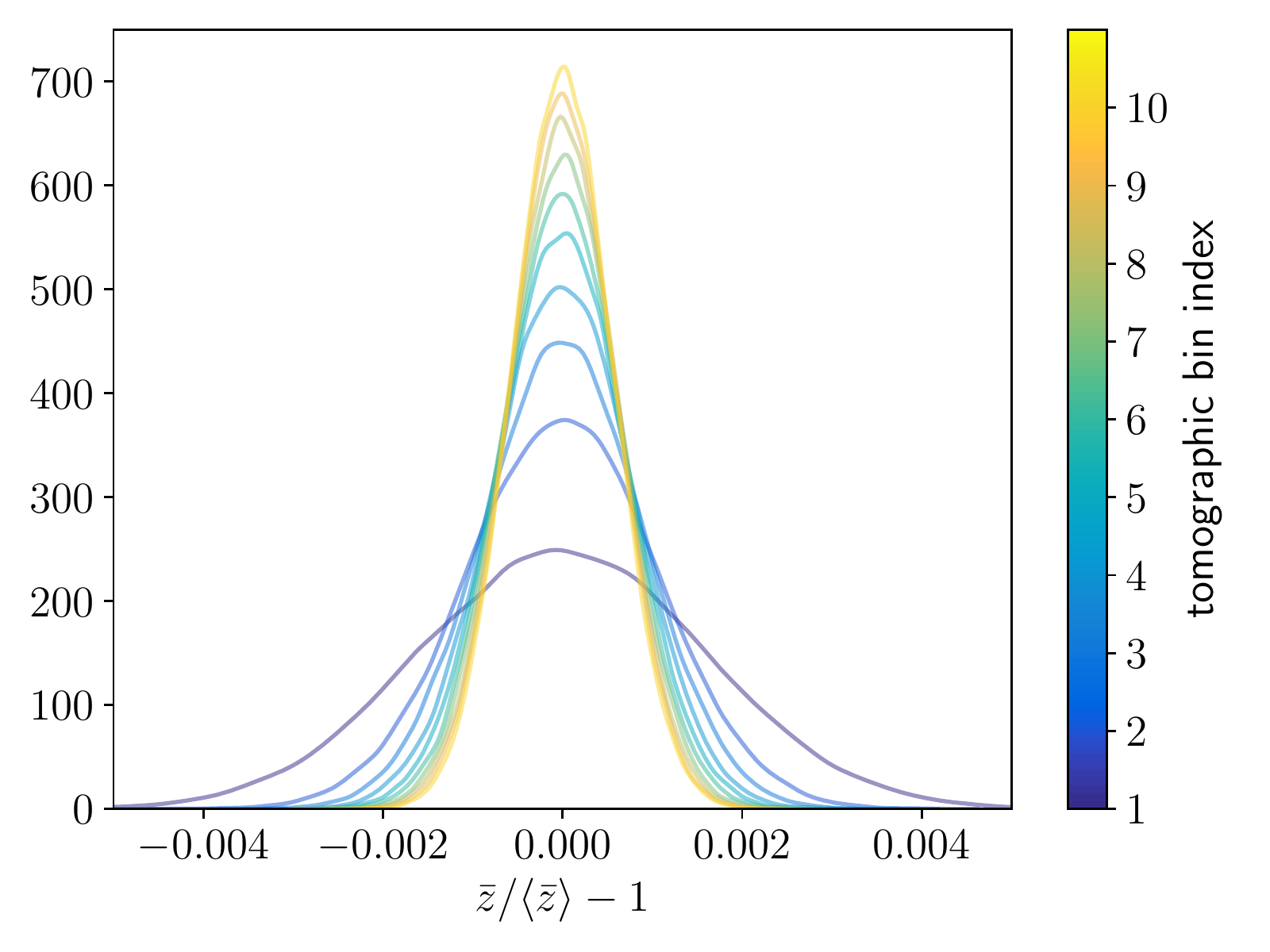}
        \includegraphics[width=0.45\textwidth]{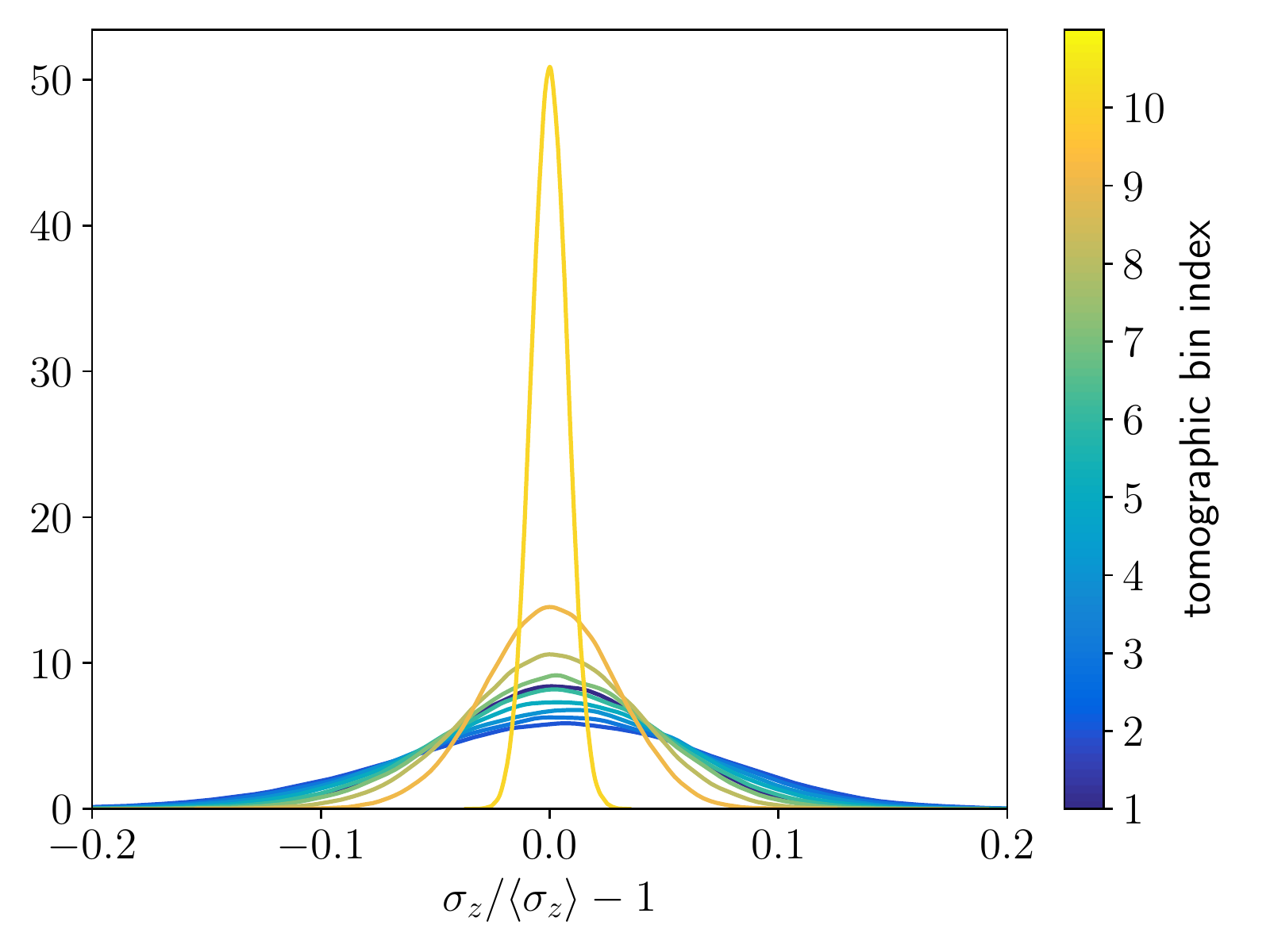}
            \includegraphics[width=0.45\textwidth]{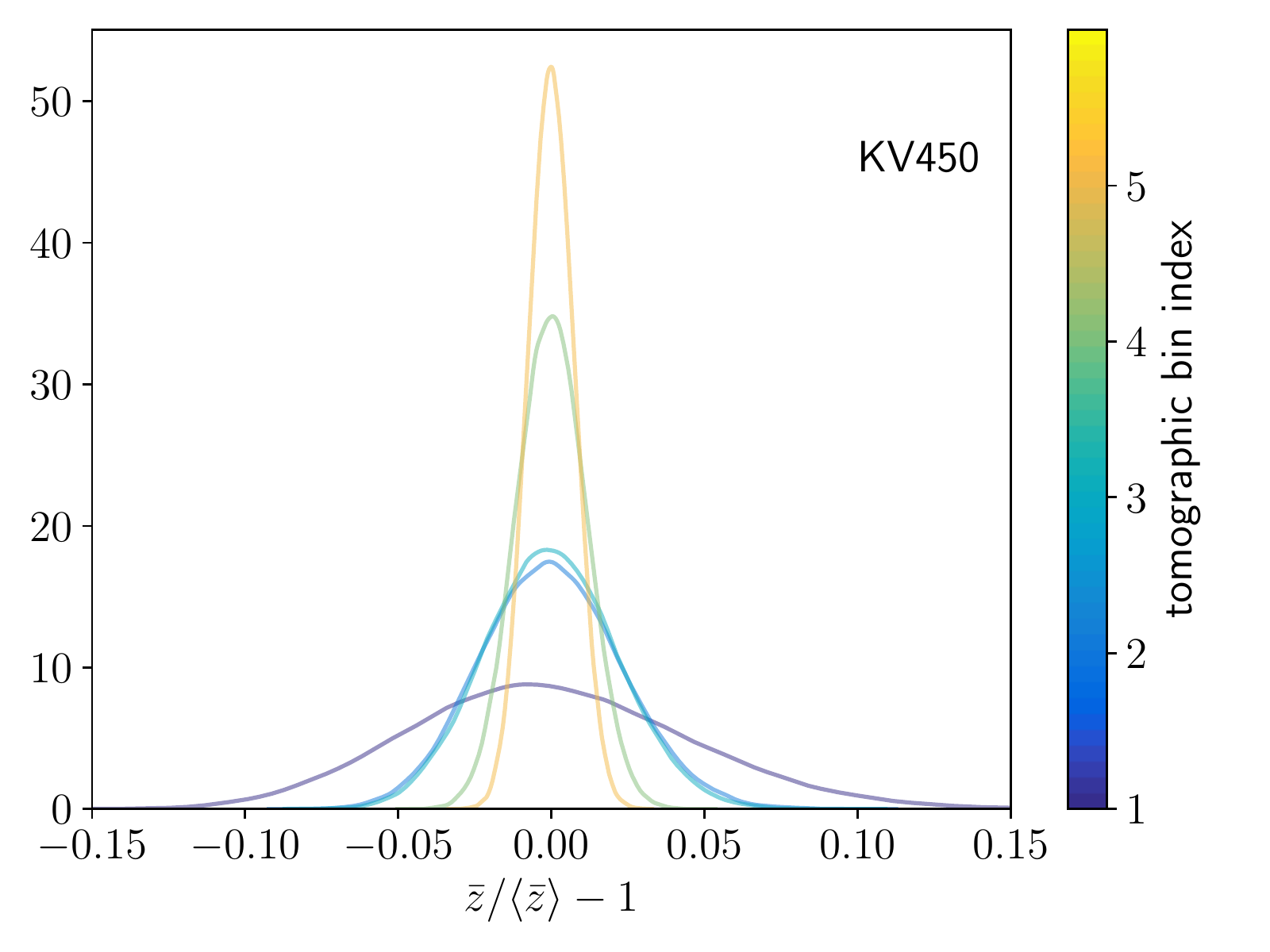}
                \includegraphics[width=0.45\textwidth]{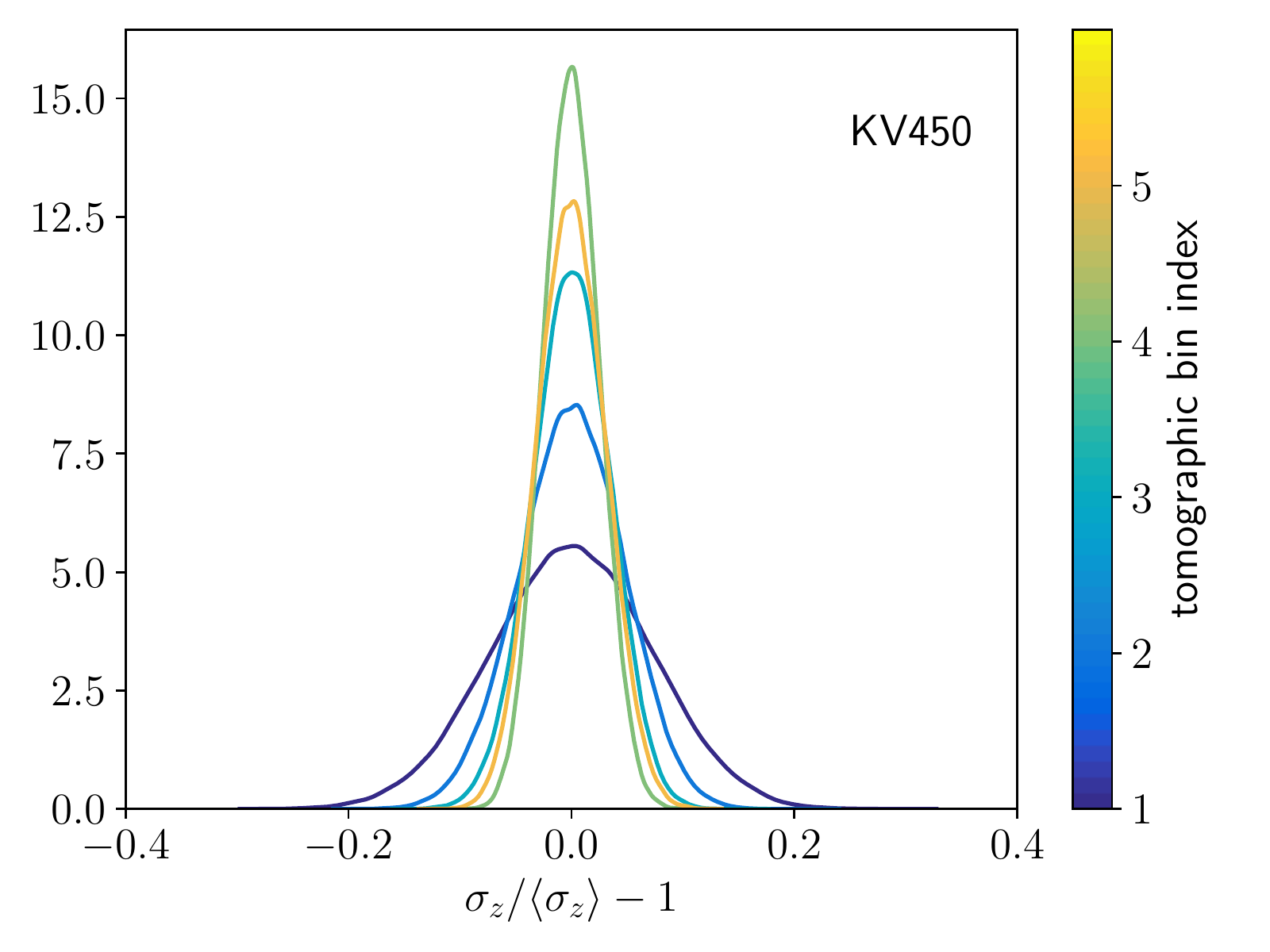}
    \caption{Distribution of the relative deviation of the mean and the variance of the SRD, $n^{(i)}_\mathrm{s}(z)$. \textbf{\textit{Top}}: For the EUCLID survey settings with realisations from the inverse of the diagonal Fisher matrix used in \Cref{fig:allowed_ecl}. \textbf{\textit{Bottom}}: For KV450 using the samples generated from the DIR bootstrap covariance.}
    \label{fig:momentskv450}
\end{figure*}

The top plots of \Cref{fig:momentskv450} show the distribution of the mean and the standard deviation and show generally good agreement with \Cref{fig:moments_euclid}, that is that the mean must be known below the per-cent level for most bins, while the standard deviation needs to be determined by roughly 10 per-cent. It should be noted that \Cref{fig:moments_euclid} considers the extreme case where we exactly look at the envelope shown in \Cref{fig:allowed_ecl}. 

Finally, the bottom two plots show the same for KV450, where we find much wider errors on mean and standard deviation, a few per-cent and a few ten per-cent respectively. The general trend, however, is the same - high redshift bins are more important than lower redshift bins.

\section{Relationships to Observables}
\label{sec:observables}
Real surveys usually do not use the angular power spectra as a final statistic. This is for example due to incomplete sky coverage, masking effects, variable depth or simply the dimensionality of the data vector. All these factors require a sufficient summary statistic. Very commonly used ones are the correlation function or band powers (or similarly pseudo-$C_\ell$). All of these are essentially linear transformations of the pure angular power spectrum $C_\ell$ and assume the following general form:
\begin{equation}
    \mathcal{O}[C_\ell] = \int\mathrm{d}\ell C_\ell W_\mathcal{O}(\ell)\;,
\end{equation}
where $ \mathcal{O}$ is some observable of interest and $W_\mathcal{O}(\ell)$ is the associated kernel defining the transformation. Again by the chain rule, the functional derivative of this new observable with respect to the SRD is readily available:
\begin{equation}
    \frac{\delta\mathcal{O}[C_\ell]}{\delta n(\chi_0)} = \int \mathrm{d}x\frac{\delta\mathcal{O}}{\delta C_\ell(x)}\frac{\delta C_\ell(x)}{\delta n(\chi_0)}\;,
\end{equation}
where we dropped all the indices for less clutter. For band powers, $\mathcal{C}_l$, this would for example assume the following form:
\begin{equation}
    \frac{\delta{\mathcal{C}_{l}}[C_\ell]}{\delta n(\chi_0)} = \frac{1}{\mathcal{N}_l}\int\mathrm{d}\ell \ell S_\ell \frac{\delta C_\ell}{\delta n (\chi_0)}\;,
\end{equation}
where $S_\ell$ is the band power response function and $N_l$ is the normalisation. For the two-point correlation function $\xi_\pm$ one finds:
\begin{equation}
    \frac{\delta \xi_\pm(\theta)}{\delta n(\chi_0)} = \frac{1}{2\pi}\int \mathrm{d}\ell \ell J_{0,4}(\ell\theta) \frac{\delta C_\ell}{\delta n (\chi_0)}\;.
\end{equation}

\section{Non-Limber $C_\ell$}
\label{sec:nonlimber}
The Limber projection used for the $C_\ell$ is not valid on large angular scales, where it must be replaced by the full expression. In full generality, for any tracers $i$ and $j$ of the matter density
\begin{equation}
    C^{ij}_\ell = \frac{2}{\pi}\int\mathrm{d}k\; k^2 I_{\ell,k,i}[n_i]I_{\ell,k,j}[n_i]\;,
\end{equation}
where the functional $I_{k,i}[n_i]$ is given by
\begin{equation}
    I_{\ell,k,i}[n_i] = \int\mathrm{d}\chi W_i[n_i] \sqrt{P_{ii}(k,\chi)}j_\ell(\chi k)\;.
\end{equation}
Here $P_{ii}$ is the auto power spectrum of the tracer $i$ and $W_i$ is its associated weight. Thus we find: 
\begin{equation}
    \frac{\delta C^{ij}_\ell}{\delta n^a} = \int \mathrm{d}k\; k^2 \left[ I_{\ell,k,i}\frac{\delta I_{\ell,k,j}}{\delta n^a}\delta^\mathrm{K}_{ja} + I_{\ell,k,j}\frac{\delta I_{\ell,k,i}}{\delta n^a}\delta^\mathrm{K}_{ia}\right]\;,
\end{equation}
where 
\begin{equation}
    \frac{\delta I_{\ell,k,i}}{\delta n^i} = \int\mathrm{d}\chi \frac{\delta W_i}{\delta n^i} \sqrt{P_{ii}(k,\chi)}j_\ell(\chi k)\;.
\end{equation}
The derivative of the weight function is calculated as before.
\vspace{-.2cm}

\section{Intrinsic Alignments}
\label{sec:intrinsics}
In this work, we have ignored intrinsic alignments (IA). Its inclusion is, however, straightforward by noting that the IA angular power spectrum is simply given by
\begin{equation}
    C^{\mathrm{II}}_\ell = \int_0^{\chi_\mathrm{H}}\frac{\mathrm{d}\chi}{\chi^2} n^{(i)}_\mathrm{s}(\chi)n^{(j)}_\mathrm{s}(\chi) P_{\mathrm{II}}\left(\frac{\ell + 0.5}{\chi},\chi\right)\;,
\end{equation}
where $P_{\mathrm{II}}$ is the IA power spectrum, which summarises the reaction of galaxy shapes to the ambient LSS on the two-point level. The functional derivative there proceeds in the same way as in \Cref{sec:photometric_clustering}. For the GI term of intrinsic alignments, one proceeds as before for cosmic shear (compare \Cref{sec:methodology}).

\label{lastpage}

\end{document}